\documentclass[a4paper,fleqn,usenatbib,useAMS]{mnras}
\usepackage[T1]{fontenc}
\usepackage{ae,aecompl}

\usepackage{graphicx}	% Including figure files
\usepackage{amsmath}	% Advanced maths commands
\usepackage{amssymb}	% Extra maths symbols

\usepackage{times}
\usepackage{epstopdf}
\usepackage{graphicx}
\usepackage{txfonts}
\usepackage{natbib}
\usepackage{array}
\title[line-of-sight contribution in substructure lensing]{Modelling the line-of-sight contribution in substructure lensing}

\author[Giulia Despali et al.]{Giulia Despali$^{1}$\thanks{E-mail:gdespali@mpa-garching.mpg.de}, Simona Vegetti$^{1}$, Simon D. M. White$^{1}$, Carlo Giocoli$^{2,3,4}$ and
\newauthor Frank C. van den Bosch$^{5}$\\
$^{1}$Max Planck Institute for Astrophysics, Karl-Schwarzschild-Strasse 1, 85740 Garching, Germany\\ 
$^2$Dipartimento di Fisica e Astronomia, Alma Mater Studiorum Universit\`{a} di Bologna, viale Berti Pichat, 6/2, 40127 Bologna, Italy\\
$^3$INAF - Osservatorio Astronomico di Bologna, via Ranzani 1, 40127, Bologna, Italy\\ 
$^4$INFN - Sezione di Bologna, viale Berti Pichat 6/2, 40127, Bologna, Italy\\
$^5$Department of Astronomy, Yale University, PO. Bo. 208101, New Haven, CT 06520-8101, USA}

\date{Accepted. Received; in original form 2017 September 1}

\pubyear{2017}

\begin{document}
\label{firstpage}
\pagerange{\pageref{firstpage}--\pageref{lastpage}}
\maketitle

\begin{abstract} 
We investigate how Einstein rings
  and magnified arcs are affected by small-mass dark-matter haloes placed along the
  line-of-sight  to  gravitational  lens systems.   By  comparing  the
  gravitational  signature  of  line-of-sight   haloes  with  that  of
  substructures within  the lensing galaxy, we  derive a mass-redshift
  relation that  allows us  to rescale  the detection  threshold (i.e.
  lowest detectable  mass) for substructures to  a detection threshold
  for  line-of-sight haloes  at  any redshift.  We  then quantify  the
  line-of-sight contribution to  the total number density  of low-mass
  objects that   can   be    detected    through   strong    gravitational
  lensing. Finally,  we assess the degeneracy  between substructures
  and line-of-sight haloes of different mass and redshift to provide a
  statistical interpretation of current and future detections, with the aim of
  distinguishing  between  CDM and  WDM.  We  find that  line-of-sight
  haloes statistically  dominate with respect to  substructures, by an
  amount that strongly depends on the source and lens redshifts, and on
  the chosen dark matter model.  Substructures represent about 30 percent of the
  total number of perturbers for low lens and source redshifts (as for
  the SLACS lenses), but less than 10 per cent for high redshift systems. We also find
  that for  data with  high enough  signal-to-noise ratio  and angular
  resolution, the non-linear  effects arising from a  double-lens-plane
  configuration are such  that one is able  to observationally recover
  the line-of-sight halo redshift with  an absolute error precision of
  0.15 at the 68 per cent confidence level.
  \end{abstract}

\section{introduction}

Strong gravitational lensing  is a powerful tool to  measure the total
projected  mass  distribution  of   structures in the Universe  from  galaxy  clusters
\citep{limousin16,meneghetti16}    to   small    sub-galactic   scales
\citep[e.g.][]{keeton03,vegetti09}.  Gravitational lensing   depends
not only on  the properties of the  system acting as a  main lens, but
also  on  the mass  distribution  integrated  along the  line-of-sight between the observer and the background source \citep{bartelmann01,bartelmann10}.
Understanding the  contribution from  the latter is, therefore,  of primary
importance  for  better  constraining the  matter  density  distribution within the Universe down to small scales.

Given the increasing resolution of the observational data, probing the line-of-sight contribution is becoming more  and more relevant and a number of recent  papers have addressed this problem,  mainly on galaxy cluster scale  systems  \citep[e.g.][]{birrer16,mccully17}. At sub-galactic scales, a significant effort has been made over the years to understand
the line-of-sight contribution to the flux-ratio anomalies observed in gravitationally lensed quasars \citep[e.g.][]{metcalf12,XuD12,XuD15}.
In particular, \citet{metcalf05} has shown that flux-ratio anomalies may
be predominantly due to low-mass dark matter haloes along the line-of-sight, as opposed to subhaloes in the host halo of the main lens; the impact of line-of-sight structures on flux ratio anomalies has then been investigated also in \citet{inoue12}, \citet{inoue16a} and \citet{inoue16b}.

The  aim of  the present paper  is  to investigate  the gravitational  lensing effect of  line-of-sight haloes on  the surface brightness
distribution of gravitationally lensed arcs  and Einstein  rings, and to quantify their
contribution to  the total number  of detectable objects. Our  goal is
also     to    provide     a     statistical    interpretation     for
current \citep{vegetti10,vegetti12,hezaveh16}   and
possible future detections of low-mass haloes. In particular,  we use simulated mock data
to explore  the relative  lensing signal  of line-of-sight  haloes and
substructures within the lens halo itself as a function of redshift, mass
and  density profile. We focus on foreground
and background  line-of-sight haloes without including the effect of subhaloes in these main haloes. These only add a minor contribution to the total line-of-sight signal.
We adopt a   general approach with the  aim of obtaining 
results that are  valid for a wide range of realistic strong lensing  
observations and we compare our results with those from \citet{li16b},  who  
carried out  a similar analysis  for a specific lensing configuration. 

In  this work,  we  show  that the  line-of-sight  contribution is  of
particular relevance when trying to distinguish between different dark
matter   models  for   four   main  reasons:   $(i)$  since   low-mass
substructures are the   surviving  cores  of  accreted progenitors  
\citep{gao04,vandenbosch05,giocoli08b},  their number  and their abundance are strongly   affected by tidal processes. In contrast   low-mass  line-of-sight  haloes
are unaffected by such processes and so provide a more robust constraint on the  mass function of dark matter haloes and so on models that  predict a strong suppression of low-mass
structures  (e.g  warm  dark  matter; WDM). Detecting even a single low-mass foreground host halo could put tight constraints on the mass of a potential WDM particle;   $(ii)$  the  number  of
detectable  line-of-sight haloes  is typically larger than the
number of detectable substructures (see Section \ref{sec_res1}), hence
failing to detect  a significant number of  small-mass structures, even
with small samples of lens  galaxies, could potentially rule  out dark
matter models that  predict a steeply rising halo  mass function (e.g
cold dark  matter; CDM);  $(iii)$ the lensing  effect of  a foreground
line-of-sight halo is larger than the lensing effect of a substructure
of  the same  mass, therefore  for a  given signal-to-noise ratio for the  lensed  images and
a given angular  resolution of the observations, line-of-sight  structures
allow  one to  probe the  dark matter  mass function  down to lower masses,   where  differences between   dark  matter   models  are larger
\citep{viel05,lovell14};  $(iv)$ finally,  the  combination of  points
$(ii)$ and $(iii)$ implies that smaller samples of lenses are required
to set constraints on  the nature of dark matter that  are as tight as
those derived when considering the substructure contribution only.

In  order to   derive   constraints on  the
(sub)halo  mass  function by comparing observations of gravitational  lensing  with
theoretical predictions, it is important to understand
the mass and density distribution of the observed structures and to adopt a common definition for all of the relevant quantities.  For example,
while  to a good approximation  isolated  dark matter  haloes
follow  NFW   density  profiles  \citep*{navarro96}  that can be characterised by their virial mass and mass-dependent concentration, subhaloes  are identified in
numerical simulations as secondary density peaks within the main halo, their density profiles are poorly represented by the NFW formula and
their  mass   is  (typically) defined   as  the  bound   mass  within   the  tidal
radius. Moreover, the  lensing signal of substructures  has often been
modelled  using Pseudo-Jaffe  profiles, which  are   truncated singular
isothermal profiles and are a poor approximation both to simulated subhalo density profiles and to the NFW formula. Discrepancies in  the mass definition and the assumed
density profiles of  (sub)haloes can result in  incorrect prediction of their lensing
properties. In  this paper we  will extensively discuss  how observed
and simulated lensing masses for  substructures and field haloes should be compared and converted into each other on the basis of
their lensing effects  in order to avoid biased  conclusions.  For
  clarity,  Table \ref{def}  lists  all  the mass definitions  adopted
  in this paper.

We separate our analysis in two parts: first, we quantify the expected
contribution of line-of-sight haloes  and substructures to the lensing
signal and  their relative  importance  for constraining the
nature of dark matter;  we then show that, once a perturbation in the
lensing signal is detected, the full lens modelling of high resolution data
can  put more stringent constraints on the position  and redshift of
the perturber  than are obtained from analytical  arguments.  In particular,
we  structure this  paper as  follows: in  Section \ref{sec_mocks}  we
describe  the analytical  models that  we  employ and  our method  for
generating mock datasets; then in Section \ref{sec_res1} we derive the
mass-redshift  relation  that  allows  us to  compare  the  effect  of
substructures with that of line-of-sight haloes at different redshifts and with different  density  profiles.   We use  these  analytical
relations  for  two  purposes:  $(i)$ to convert  the  lowest  detectable
substructure mass to a lowest detectable field halo mass, as a function
of  redshift; and $(ii)$   to  correctly
integrate the  line-of-sight mass  function by  considering only
those haloes that  would have a detectable lensing  effect. In Section
\ref{sec_res2}, we model  our mock datasets using the  lensing code of
\citet{vegetti09} to  quantify the  degeneracies in  the mass-redshift
space and  to test the limits of  the analytical approach derived  in the
previous section. This allows us to statistically interpret individual
detections from observations and to quantify the  probability that these
arise from a line-of-sight halo.  Finally, in Section \ref{sec_cloncl}
we conclude by summarizing our results.

\section{Mock Data} \label{sec_mocks}

\begin{table*} \centering
\caption{Summary of the main mass definitions and  notations used
  throughout this paper. In general, the superscript indicates the assumed
  density profile. \label{def}}
\begin{tabular}{cc} \hline 

\multicolumn{2}{c}{Summary of mass definitions} \\ 
\hline 
$M_{\rm tot}^{\rm PJ}$ & Total mass of PJ profile (equation \ref{pjmass})\\\\
$M_{\rm low}$ & Detection threshold (i.e. lowest detectable mass)
                 derived  from observations, under the assumption that
                 perturbers\\ 
& are PJ subhaloes located on the plane of the host lens; for our purposes, it can be considered equivalent to 
$M_{\rm tot}^{\rm PJ}$\\ \\
$M_{\rm vir}^{\rm NFW}$ & Virial mass of NFW haloes, adopted for
                       line-of-sight haloes, and where the virial
                       overdensity is defined \\&following \citet{bryan98} \\ \\
$M_{\rm SUB}$ & SUBFIND subhalo mass \\ \\
$M_{\rm sub}^{\rm NFW}$ & Virial mass of the NFW profile that best fits the deflection angle of simulated subhaloes\\
\hline
\end{tabular}

\end{table*}

In this section, we describe the mock gravitational lenses used for our simulations.

\subsection{Input lens and source models}

In order  to test  the general  validity of  our results,  we consider
several mock data  sets. These are characterized  by different angular
resolutions,  signal-to-noise ratios,  background source  morphologies
and lens-source alignments, as well as by perturbers located at different
redshifts.  More details on the lens systems considered here are given
in Table~\ref{tab_lenses}.  

In  our simplest model,  the source  has a
Gaussian light profile and, to avoid any influence from asymmetry, the
main lens has a singular isothermal sphere (SIS) mass profile with no
external shear and the lens and source are perfectly
aligned (complete Einstein ring).  We use this toy model as a reference, in particular, to compare
our results with those of \citet{li16b}.  We then modify this model by
adding ellipticity and external shear in order to systematically test
the effect of  these components.  The other lens models are based on
real observations; this means that  the lens models include
both  ellipticity  and   external  shear,  and  the  source models  are  not
regular, but are based on actual lensed galaxy surface brightness distributions. 
In  particular, we base our  mock data on: $(i)$  two systems
from the SLACS  survey \citep{bolton06}, which have  already been used
for the analysis of substructure by \citet{vegetti10,vegetti14} and \citet{despali17b};
$(ii)$  three  systems are taken from a sample  of  $z\sim2.5$ lensed  Lyman  alpha
emitting galaxies \citep{shu16a,shu16bis}; and $(iii)$ one system from the
SHARP survey \citep{lagattuta12}
that \citet{vegetti12}  used to detect a 
$1.9\times 10^8$~M$_\odot$ substructure.

\begin{figure*}
\begin{center}
\includegraphics[width=0.24\hsize]{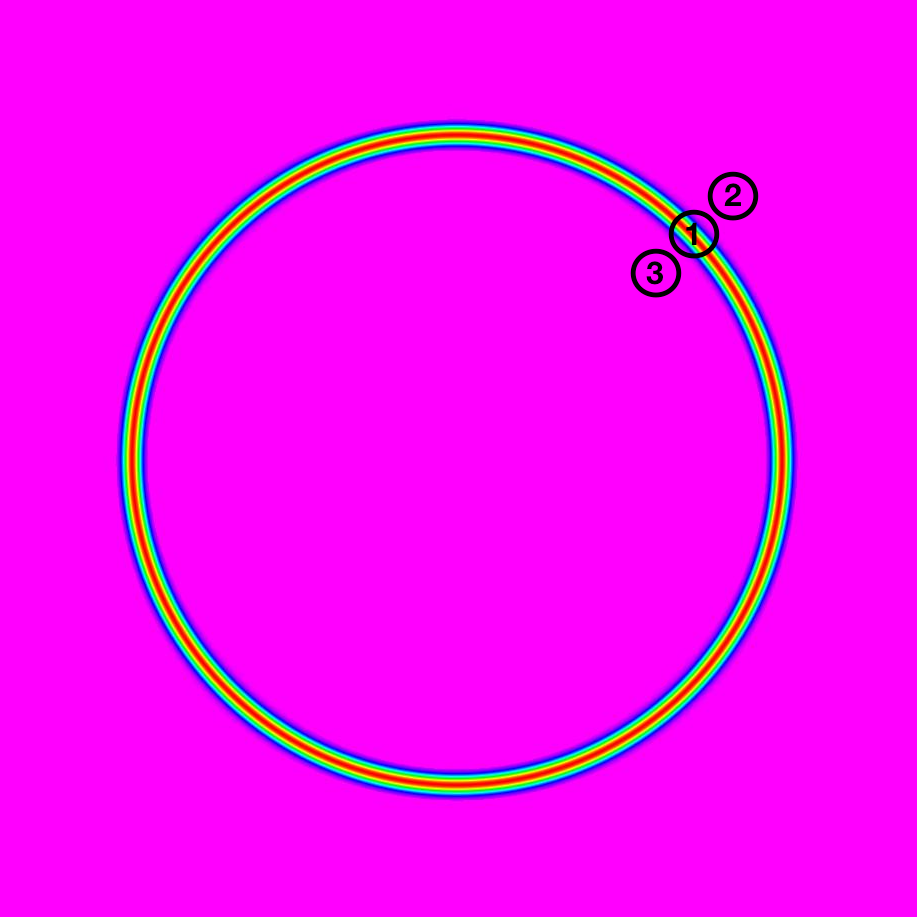}
\includegraphics[width=0.24\hsize]{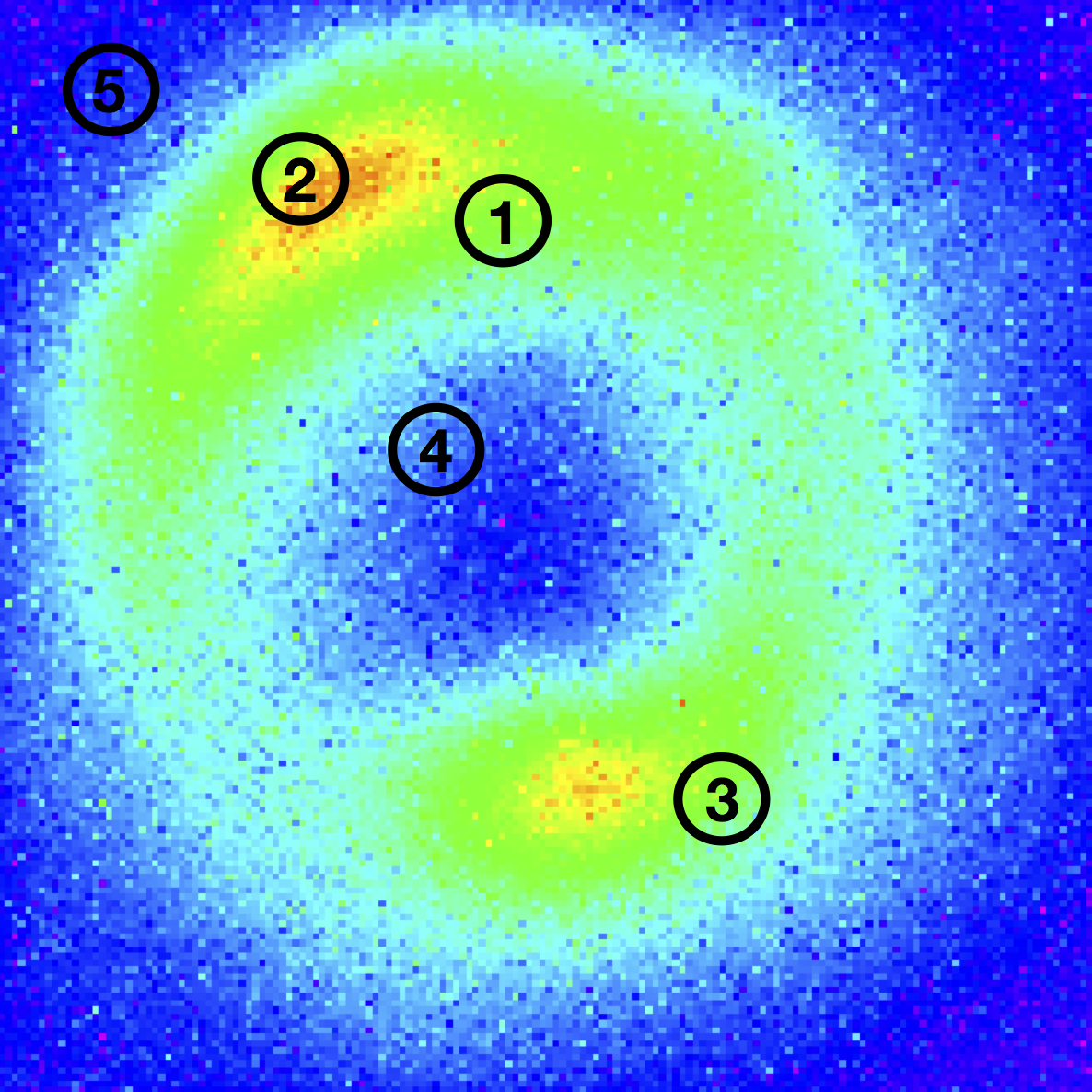} 
\includegraphics[width=0.24\hsize]{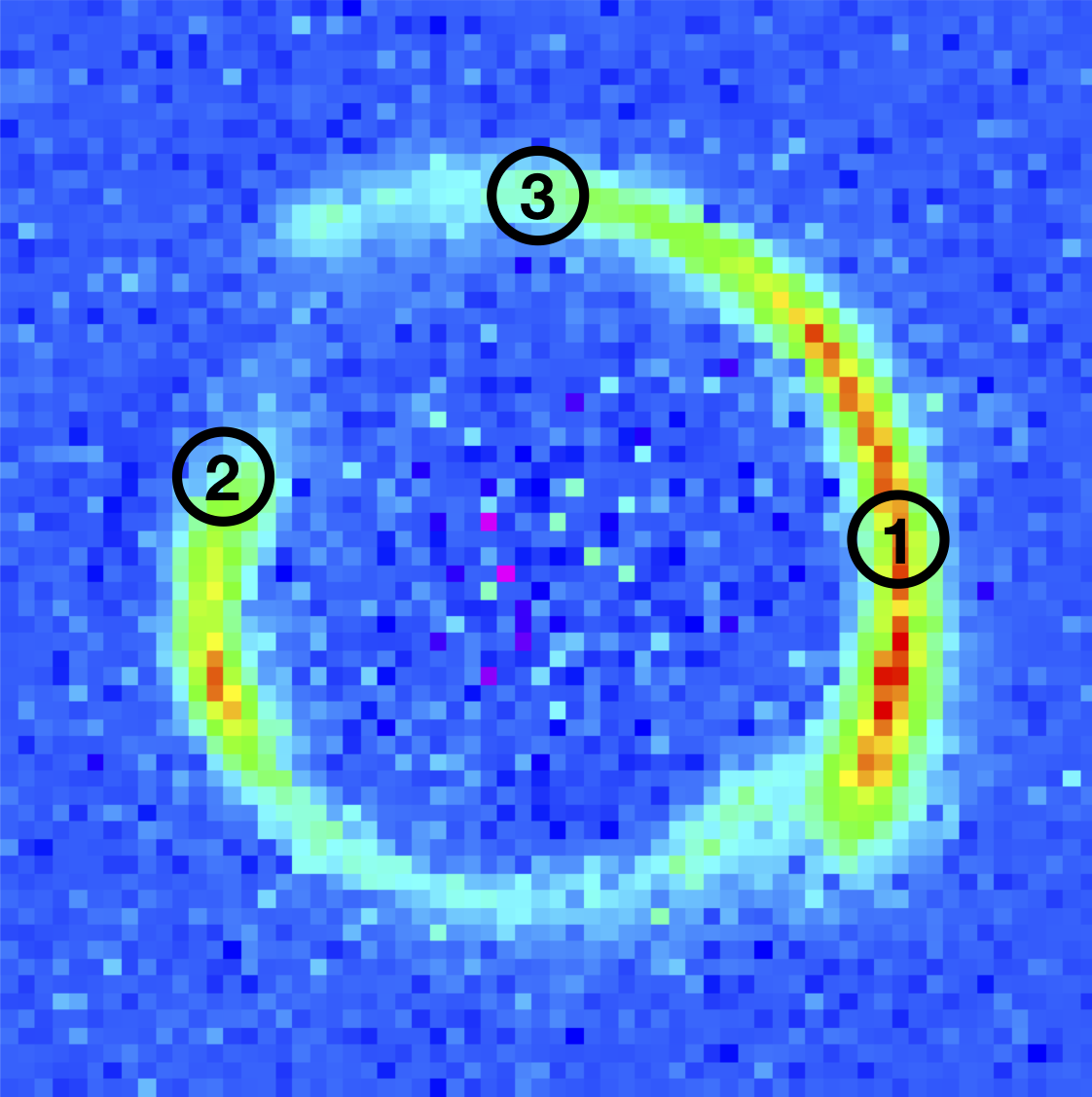}
\includegraphics[width=0.24\hsize]{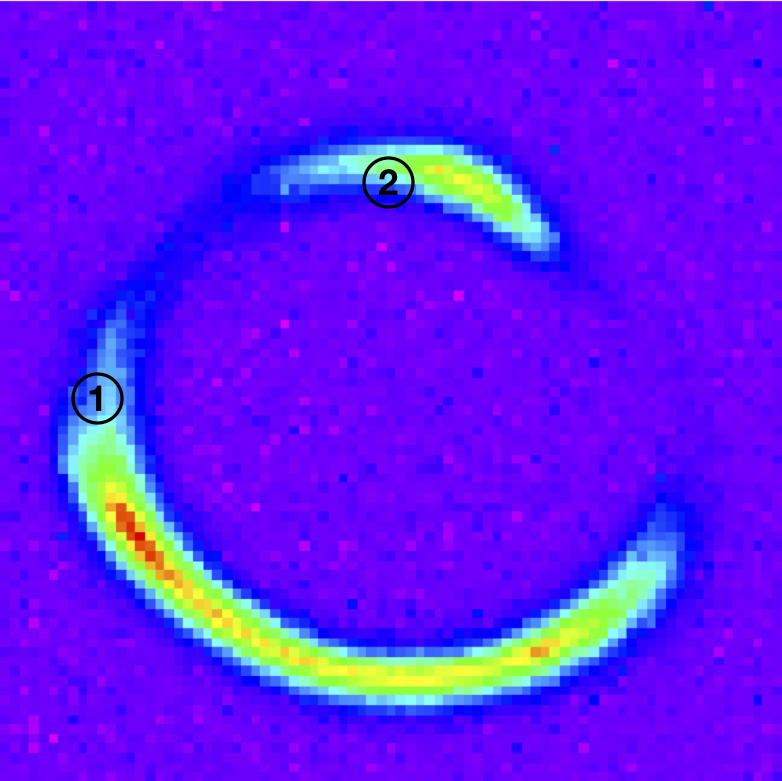}\\
\includegraphics[width=0.24\hsize]{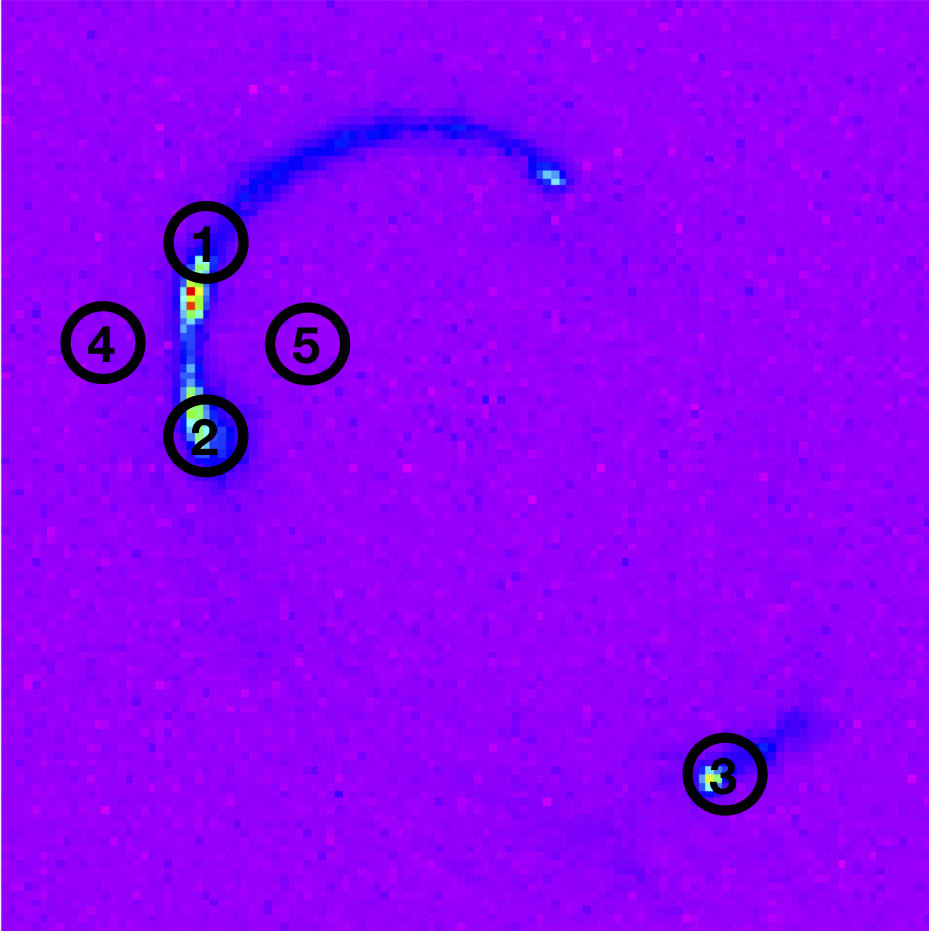}
\includegraphics[width=0.24\hsize]{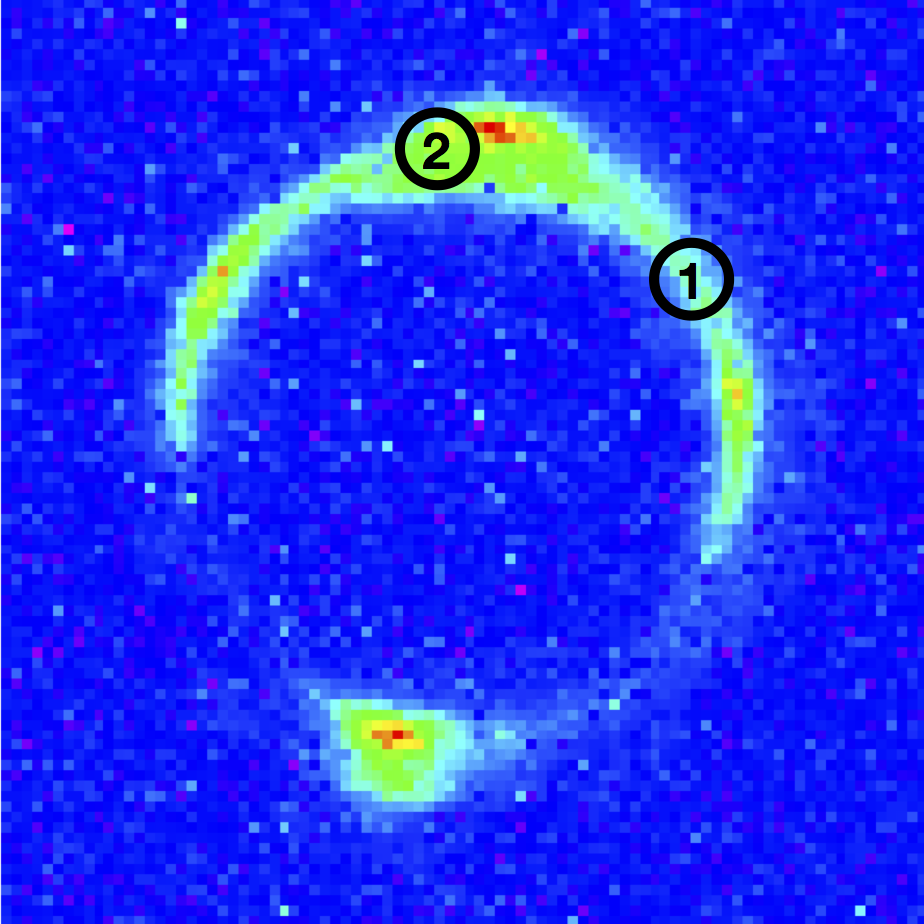}
\includegraphics[width=0.24\hsize]{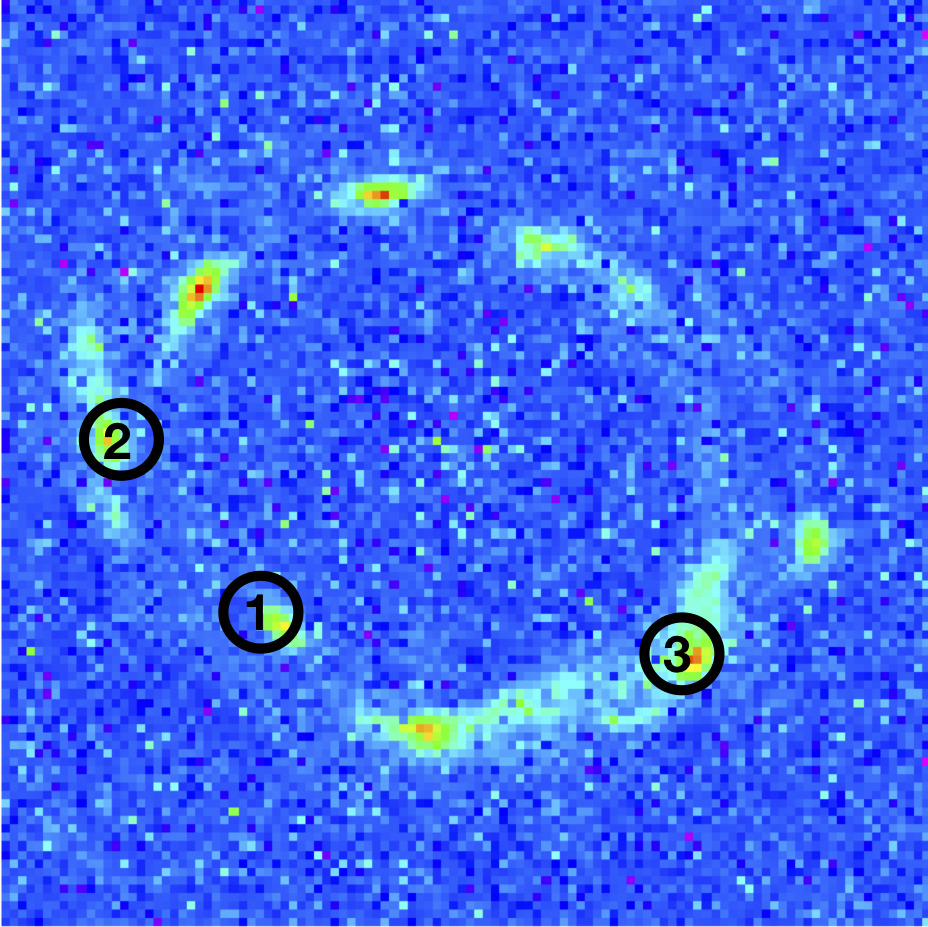}
\end{center}
\caption{The gravitational lens systems and the projected positions for the perturbers. For each lens system we create several mock datasets with perturbers located at the projected positions indicated by the
circles. In the first row, we have a SIS analytical lens model with a Gaussian source, one mock dataset based on JVAS B1938+666 (SHARP; Keck adaptive optics) and two on the SLACS systems J0252+0039 and J0946+1006 ({\it HST}); the mock datasets on the second row are based on the {\it HST}-observed BOSS lenses J0918+5104, J1110+3649 and J1226+5457. The lens properties are listed in Table \ref{tab_lenses}. 
\label{lenses} }
\end{figure*}

For each lens  system, we  consider a so-called \emph{smooth}
model, that is, without  any substructure or  line-of-sight  halo, and
several \emph{perturbed} models, where substructures and line-of-sight
haloes  with  different masses,  redshifts  and  density profiles  are
included  (see Section  \ref{sec_2planes}  for  details). Mock  images
that were created   using  the    smooth    models   alone    are   shown    in
Fig. \ref{lenses}.  In the  next sections we  provide further details on the properties of the perturbers.

\begin{table} \centering
\caption{Properties of the gravitational lens systems considered in this paper (see also Figure \ref{lenses}). For
 the first lens we use a Gaussian source with a SIS/SIE analytical lens model with different combinations of axial ratio ($q$) and external shear  strength ($\Gamma$), while in the other cases
 both the source and the SIE lens models are based on the lens modelling of real observations. For each lens system we also quote the lens and source redshifts $z_{l}$ and $z_{s}$ and the number of pixels $N_{\rm pix}$ on the plane of the lens used to quantify the lensing effect of different perturbers. 
\label{tab_lenses}}
\begin{tabular}{lccccc} \hline 
Name & \multicolumn{5}{c}{Lens
models} \\ & $z_{l}$ & $z_{s}$  & $q$ & $\Gamma$ & $\sqrt N_{\rm pix}$\\

\hline 
Analytical SIS/SIE & 0.20 & 1.00 & 1/0.8/0.6 & 0/0.26/0.3 & 512\\
JVAS B1938+666 & 0.88 & 2.06   & 0.82 & 0.04 & 165\\ 
SLACS J0252+0039 &0.28 & 0.98 &0.94 &0.01 & 64\\ 
SLACS J0946+1006 &0.22 & 0.61  & 0.96 & 0.05 & 81\\ 
BOSS J0918+5104 & 0.58 & 2.40  & 0.65 & 0.25 & 120\\ 
BOSS J1110+3649 & 0.73 & 2.50  & 0.86 & 0.02 & 90\\ 
BOSS J1226+5457 & 0.50 & 2.60   & 0.97 & 0.15 & 110\\ 
\hline
\end{tabular}
\end{table}

\subsection{Inclusion of haloes along the line-of-sight}\label{sec_2planes}

\begin{figure}
\includegraphics[width=\hsize]{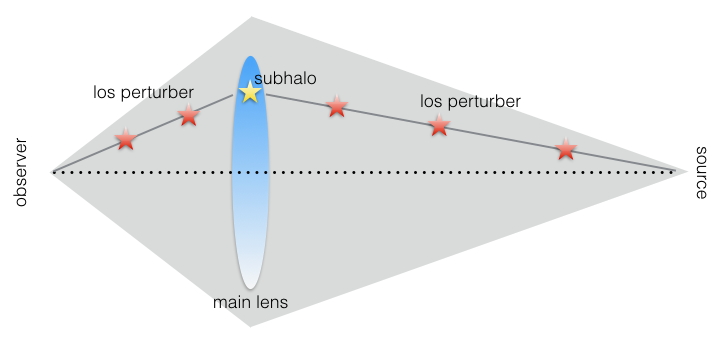}
\caption {A simple sketch of the method we used to
create our mock data; subhaloes and line-of-sight haloes are placed so that their lensing effect lies in the same
projected position on the plane of the main lens; the grey region gives an example of the line-of-sight volume that is taken into account.\label{sketch}}
\end{figure}

In  order  to  include  only   those  line-of-sight  haloes  that  can
effectively perturb  the lensed  images, we consider a line-of-sight
volume that is a double cone with a base of 1.5 times the Einstein radius of
the main lens  (see Figure \ref{sketch}). Within this  cone, we sample
the whole  redshift range between the  observer and the source, thus
considering  both foreground  and  background perturbers.  The line-of-sight
haloes  are modelled  as  NFW  profiles,
while for the substructures  we  consider both NFW and Pseudo-Jaffe (PJ)
profiles; the latter are often   used   to   model   real      datasets
\citep[e.g.][]{dalal02,  vegetti14, hezaveh16}. Moreover, it  is well known that isolated  dark matter haloes and subhaloes  do not have  the same  profiles, since the  latter have been subjected to  tidal interactions with the main  halo after infall and  may have   been   stripped    of   significant   amounts   of   mass \citep{hayashi03,giocoli08b}.  Here, we are interested in (sub)haloes that do not have a bright stellar component and we assume  the
highest     possible  subhalo  mass   to   be
$\simeq   10^{10}$~M$_{\odot}$. 
 The  minimum  mass  is  
chosen in  such  a  way as to  include
line-of-sight haloes that are relevant for substructure detections, which in this case is 
PJ-like haloes down to $M_{\rm tot}^{\rm PJ}=10^{6}$~M$_{\odot}$. Both limits are set in
terms of the total mass of the PJ profile in the plane of the host lens.  Following  the
conversion between the PJ and NFW profile masses at different
redshift (see Section \ref{sec_comp_prof}),   we   set   the   relevant  range  for  the NFW profile masses to lie between $10^{5}$ and $10^{11}$~M$_{\odot}$.  

In  the perturbed  models, substructures have  projected positions  as
marked by  the numbered circles in  Fig.~\ref{lenses}. As we  want to
perform a  one-to-one comparison between  the local lensing  effect of
the two  different populations, the  2D position of  the line-of-sight
haloes is corrected  with redshift in such a way  that they affect the
lensed images  at the same  position as substructures within the lens would.  This means
that  the   line-of-sight  halo   should  always   lie  on   the  same
\emph{line-of-sight},   as  sketched   in  Fig.~\ref{sketch}.   In
particular, we  use the factor  $\beta$ (see Section  \ref{sec_res1} and
equation \ref{beta} for a definition) to rescale the position of any
perturber behind the lens.  For  each \emph{perturbed} model we only
consider the presence of one perturber at a time; this is justified by
the fact  that we are  interested in quantifying the  relative lensing
effect  of substructures  and line-of-sight  haloes rather  than their
global effect on the data.

We now summarize the main features of the mass
profiles considered here, and the basic equations  used to calculate  their deflection
angles.

\subsection{NFW profile}

The NFW density profile is defined as,
\begin{equation}
\rho(r)=\dfrac{\rho_{s}}{\frac{r}{r_{s}}\left(1+\frac{r}{r_{s}}\right)^{2}},
\end{equation}
where $\rho(r)$ is the density as a function of radius $r$, the scale-radius is given by $r_s$, and $\rho_s$ is the density normalization. The NFW profile 
can also be  defined   in  terms  of  the  halo   virial  mass  $M_{\rm vir}$
\citep[i.e.  the  mass   within  the  radius  that   encloses  a  virial
 overdensity $\Delta_{\rm vir}$, defined  following][]{bryan98}, 
and a concentration related to the scale radius through  $r_{s}=r_{\rm vir}/c_{\rm vir}$.

Throughout this paper we adopt the concentration-mass relation by  \citet{duffy08} for the to relate halo concentration to virial mass and redshift, and we ignore the presence of scatter,  meaning  that  we  assign   a  deterministic  value  of  the
concentration  for each combination  of mass  and redshift.  In
Appendix  \ref{sec_appendix_a},  we  demonstrate  that  for  the  main
purposes of this paper, a  different choice for the mass-concentration
relation  or  allowing  for some  scatter around  the  mean  value introduces only second  order effects. When modeling subhaloes as having NFW density profiles, we assume that they follow the same concentration-mass relation as host haloes. We discuss the validity and the implications  
of this assumption in Section \ref{sec_effective}.  

Starting  from the dimensionless form
of the  lens equation, where $x=\theta/\theta_{s}$  (with $\theta_{s}$
being the angular-scale associated with  $r_{s}$), the deflection angle can be
written as,
\begin{equation} 
\alpha(x)=\dfrac{4k_{s}}{x}h(x),
\end{equation}
where
\begin{equation} h(x)=\ln\dfrac{x}{2}+
\begin{cases} \dfrac{2}{\sqrt{x^2-1}}\arctan\sqrt{\dfrac{x-1}{x+1}} &
\text{if} (x>1) \\ \dfrac{2}{\sqrt{1-x^2}}\mathrm{arctanh}\sqrt{\dfrac{1-x}{x+1}}
& \text{if} (x<1) \\ 1 & \text{if} (x=1)
\end{cases}
\end{equation}
and
\begin{equation} k_{s}=\dfrac{\rho_{s}r_{s}}{\Sigma_{c}} ,
\qquad \Sigma_{c}=\dfrac{c^{2}D_{s}}{4\pi G D_{l}D_{ls}}.
\end{equation}
Here $\Sigma_{c}$  is the critical  surface mass density, and $D_{l}$,
$D_{s}$  and $D_{ls}$  are  the angular  diameter  distances from  the
observer to the lens, the observer to the source, and the lens
to the source, respectively.

\subsection{PJ profile}
\label{secpjprof}
The Pseudo-Jaffe profile is defined as
\begin{equation}
\rho(r)=\dfrac{\rho_{0}r_{t}^{4}}{r^{2}(r^{2}+r_{t}^{2})},
\end{equation}
and corresponds to the convergence
\begin{equation}
\kappa(R)=\kappa_{0}r_{t}\left[R^{-1}-(R^{2}+r_{t}^{2})^{-1/2}\right],
\end{equation}
where $r_{t}$ is the   truncation    radius, $\rho_0$ is the density normalization, and the convergence normalisation is $\kappa_{0}=\pi\rho_{0}r_{t}/\Sigma_{c}$.

The profile deflection angle  as a function of  the substructure projected  position is expressed as
\begin{equation}
\alpha(R) = \alpha_{0}\dfrac{r_{t}+R-\sqrt{r_{t}^{2}+R^{2}}}{R}, 
\end{equation}
where $\alpha_{0}=2r_{t}\kappa_{0}D_{s}/(D_{l}D_{ls})$. Then the total mass - obtained by integrating out to infinity - can be written as
\begin{equation} 
M_{\rm tot}^{\rm PJ}=2\pi\Sigma_{c}r_{t}^{2}\kappa_{0}. \label{pjmass}
\end{equation}
Generally, the truncation radius is assumed to  be well approximated by the substructure tidal radius
\begin{equation} 
r_{t}\simeq r_{\rm tidal}=r\left(\dfrac{M_{\rm tot}^{\rm PJ}}{\xi M(<r)}\right)^{1/3}, \label{r_tidal}
\end{equation}
which, for a singular isothermal host lens, reduces to
\begin{equation}
r_{t}=r\sqrt{\dfrac{\pi \kappa_{0}}{2\xi \kappa_{0,{\rm L}}}}. 
\end{equation}
Here, the impact parameter $\xi$ depends  on  the  assumptions  made  on the  satellite  orbit  (it  is
typically  set equal  to 3  for  the assumption  of circular  orbits),
$\kappa_{0,{\rm L}}$ is  the convergence normalization of the  main lens 
and  $M(<r)$ is the mass of the host halo enclosed within the radius, r, which is equal to the distance of the subhalo from the centre of the host halo.  Thus, the truncation of  the profile depends on the redshift
(via  $\Sigma_{c}$) and  mass  of the  host lens  galaxy,  and its  3D
position  relative to  the  centre of  the host.  However,  in a  real
situation, this  distance is  not known  and one  can only  measure the
two-dimensional   distance  $R$   projected  on  the  plane   of  the
host.  Therefore,  one  generally  assumes that  the  substructure  is
located on  the plane of  the host  lens, that is, $r=R$.  Throughout this
paper, when  we refer to a perturber with a PJ profile, we  always make use  of this
assumption.  We discuss this issue and its implications in more detail
in Appendix \ref{sec_appendix_c}.

Finally, as the normalization of the  PJ profile for a sub-halo depends
on  the mass  of the  main halo  it is  embedded in,  it would  not be
meaningful to define a virial mass  or virial radius for this profile in the same way as is the case for the NFW profile. In Section \ref{sec_comp_prof} we  investigate how to
compare the  NFW equivalent virial mass  and the PJ total  mass on the
basis of their lensing effects.

\section{A model for line-of-sight haloes} \label{sec_res1}

The  aim  of  this  section  is to  understand  how  to  quantify  the
line-of-sight contribution  to the total number  of detectable objects
(i.e.    substructures  plus   line-of-sight  haloes)   for  different
lens-source  redshift   configurations.  

Both  contributions   can  be
quantified by integrating  the relative mass function  from the lowest
detectable   mass    to   the   highest   possible    dark   (sub)halo
mass.   \citet{vegetti14}   have   defined   the   lowest   detectable
substructure mass as the mass that  can be detected with a statistical
significance of 10$\sigma$.   We refer to their paper  for a detailed
discussion on how  this mass is determined. What is  important to know
for the purpose of this paper  is that this detection limit is derived
for substructures with  a PJ profile located on the  plane of the host
lens.   In   principle,   following    the   same   approach   as   used by 
\citet{vegetti14}, one could derive the detection limit for any choice
of the perturber mass-density profile  and redshift.  However, this is can be 
computationally expensive.  The aim of  this section is therefore to
derive simple  analytic relations  that allow one  to rescale  a given
detection  limit,   by  comparing  the  relative   lensing  effect  of
substructures  and line-of-sight  haloes with different redshift and
mass-density profiles. Given a  certain detection limit
$M_{\rm low}(z=z_{\rm L})$  for substructures  in the  lens, calculated under the
assumption that the perturber is a PJ subhalo in the plane of the host lens, our aim here is to
derive analytical relations to convert this mass into an effective $M_{\rm low}(z)$ that 
we can use for the integration limit for the (sub)halo mass function.

First, we investigate how  the detection limit
would change with redshift for perturbers with a NFW profile and then,
to reproduce what  is done in the modelling of  actual data, we assume
that this limit has been derived for substructures with a PJ
profile. In particular, we investigate how to compare the lensing effect of
these two density profiles and we discuss whether they are a good
model for the (sub)haloes.

\subsection{Lensing effect} 

Most  previous  studies  on  the effect  of  line-of-sight  haloes  on
gravitationally lensed  images have focused mainly  on multiply-imaged
quasars   \citep[e.g.][]{chen03,metcalf05,XuD12}.  Therefore,   the  relative
gravitational  lensing effect of  substructures and  line-of-sight haloes  has
been   quantified  in   terms  of   local  changes   to  the   lensing
magnification. In this  paper, we focus instead on  Einstein rings and
magnified arcs.  

As demonstrated by  \citet{koopmans05}, perturbations
to  the   lensing  potential  locally  affect   the  observed  surface
brightness distribution with  a strength that can be  expressed as the
inner  product  of  the  gradient of  the  background  source  surface
brightness distribution ($\nabla \mathbf{s}$;  evaluated in the source
plane) dotted with  the gradient of the potential  perturbation due to
(sub)structures ($\nabla \delta\mathbf{\psi}$;  evaluated in the image
plane), such that,
$\nabla      \mathbf{I}      =      -\nabla      \mathbf{s}      \cdot
\nabla\delta\boldsymbol{\psi}$.
Since the (sub)structure deflection angle  is related to its potential
as $\delta\boldsymbol{\alpha} =  \nabla\delta\boldsymbol{\psi}$, for a
given  background  source  brightness distribution,  we  quantify  the
relative  gravitational  effect  of  substructures  and  line-of-sight
haloes  in terms  of their  deflection  angles. In  particular, for  a
substructure of  a given mass  and projected position relative  to the
main lensing galaxy, at each redshift $0\leq z \leq z_{\rm S}$ we look for
the line-of-sight halo mass that, at the same
projected position, minimizes the following deflection angle residuals
\begin{equation} 
\mathrm{d}\alpha = \left(\dfrac{1}{N_{\rm pix}}\sum_{i=1}^{N_{\rm pix}} (\Delta\boldsymbol{\alpha}_{\rm LOS}-\Delta\boldsymbol{\alpha}_{\rm sub})^{2}\right)^{1/2},
\label{def_res}
\end{equation}
where  $\Delta\boldsymbol{\alpha}_{i}$   is  the  difference   in  the
deflection angle  between the perturbed  and the smooth model, and d$\alpha$ is the average over the pixels on the lens plane. The number of pixels, $N_{\rm pix}$, is kept constant for each mock system.

In the simple  case of  two  lenses at  the same  redshift,  both the  lensing
potential and the deflection angle can be written as the linear sum of the
individual contributions  of the two  lenses, and the lens  equation is
written as,
\begin{equation} 
\mathbf{u}
=\mathbf{x}-\left[\boldsymbol{\alpha_{1}}(\mathbf{x})+\boldsymbol{\alpha_{2}}(\mathbf
{x})\right], \label{def_res2}
\end{equation}
where \textbf{u} and \textbf{x} are the true and observed positions of the source, respectively, and $\alpha_{i}$(\textbf {x}) is the deflection angle of the $i$-th lens at the \textbf{x} position on the lens plane.
When two lenses are sufficiently separated along the line-of-sight for their
caustics to be distinct, a  recursive lens equation is required instead
\citep{schneider92},
\begin{equation} 
 \mathbf{u}
=\mathbf{x}-\boldsymbol{\alpha_{1}}(\mathbf{x})-\boldsymbol{\alpha_{2}}\left[\mathbf
{x}-\beta\boldsymbol{\alpha_{1}}(\mathbf{x})\right],
\label{2planes}
\end{equation}
where the factor
\begin{equation} 
\beta=\dfrac{D_{12}D_{os}}{D_{o2}D_{1s}}
\label{beta}
\end{equation}
encodes  the  redshift  difference  in terms  of  the distance  ratio  for
$z_{2}\geq z_{1}$;  $\beta$ vanishes if  the two lenses have  the same
redshift and approaches  unity for redshifts close to  the observer or
the  source.    The  squared  brackets  following  $\boldsymbol\alpha_{2}$  in
 equation  (\ref{2planes})  contain its arguments and indicates  that  the  position   at  which
 $\boldsymbol \alpha_{2}$ is  evaluated depends  on the  deflection angle  of the foreground lens.  
   
When comparing  the lensing  effect of  a given
substructure with  line-of-sight haloes via  equation (\ref{def_res}),
we  first  order  the  lenses  in redshift  and  then  apply  equation
(\ref{2planes}).  As shown by  \citet{mccully17}, since the deflection
angle  of the  foreground lens  enters the
argument  of the  deflection angle  of the  background lens,  
non-linear lensing effects are introduced when  the mass of the former is large
enough. However, the masses of our perturbers are much smaller than the  
mass of the main lens, by 3 to 7 orders  of magnitude,  and thus  when the  perturber is  in the
foreground its  effect on the main  lens is small, while  the opposite
holds when the perturber is in the background and its deflection angle
is influenced by the presence of the main lens.

The exact width of the image plane
varies from one  mock dataset to another, and  ranges from 1.6 arcsec
for the  SHARP lens, to  3--4 arcsec for the  SLACS and HST  lenses, to about 8 arcsec  for the idealized model  using a SIS
lens and a Gaussian source.

\subsection{Deflection angles at different redshifts}

\label{defangdz}
Before  investigating   the  effects  due  to   the  double-lens-plane
coupling, we  want to study how  the lensing properties of haloes with an NFW profile
evolve as  a function of  redshift. To this end,  we choose a lens with a NFW profile that has a  virial mass   of  $M_{\rm ref}=10^{7}$~M$_{\odot}$   and  a   redshift  of
$z_{\rm ref}=0.2$  as a  reference point.  Then, at  each redshift  in the
considered range we  find the virial mass of the NFW  lens that minimizes 
the value of
\begin{equation}\label{dr}
D_r = \sqrt{\dfrac{\sum_i(\alpha_i - \alpha_{i,ref})^2}{ \sum_i(\alpha_{i,ref})^2}}
\end{equation}
where $\boldsymbol\alpha$ is the deflection angle of a NFW lens at a certain redshift $0<z<z_{\rm S}$ and $\boldsymbol\alpha_{ref}$ is the deflection angle of the reference case.
Note that here we force the NFW haloes at each redshift to follow our reference concentration-mass relation. Also, here we only compare the deflection angles of individual NFW profiles, without the contribution of a main halo. We will introduce a host galaxy and the double-plane lensing in the next section.
For this analytical comparison, we calculate the deflection angle in a region of 5~arcsec$^2$, which includes a total of $512^2$ pixels, and we calculate the average relative difference in this region. This is large enough to enclose the scales that are relevant for the lensing signal of the masses considered here, since it is substantially larger than the region in which the deflection angle is close to its maximum value.
  
\begin{figure}
\includegraphics[width=1
\hsize]{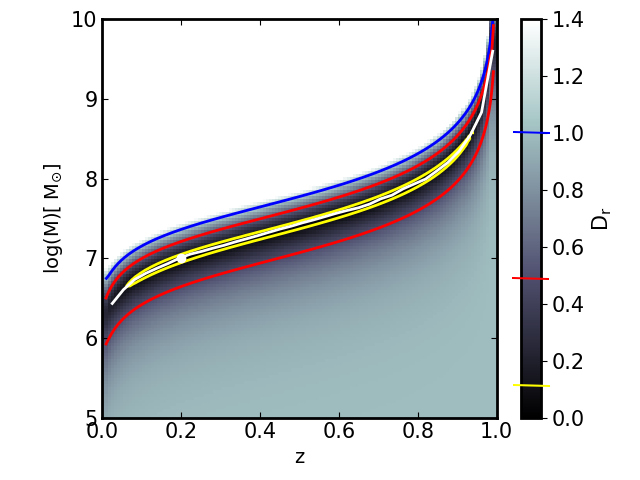}
\caption{A measure of relative difference in the deflection angle using equation (\ref{dr})
in the $z$--$\log(M)$ plane, for NFW haloes at different redshifts. The white dot marks the reference case used for the comparison: the gray scale shows the value of $D_{r}$ of all other combinations with respect to the reference case ($M_{\rm ref}=10^{7}M_{\odot}$ at $z_{\rm ref}=0.2$), while the white solid line shows the minimum of the residuals at each redshift. The coloured contours enclose the points for which the value of $D_{r}$ is within 0.1, 0.5 and 1 - as marked on the color bar. 
\label{deg_nfw}}
\end{figure}

The results  are  presented  in  Fig.~\ref{deg_nfw}.  As  expected  from 
geometrical  arguments, for  a  given  mass   and  a fixed  source
  redshift $z_{\rm S}$,  the deflection  angle decreases  with increasing
redshift. Therefore,  given a certain ($z_{\rm  ref}$, $M_{\rm ref}$), a
similar  deflection  angle  may  result from  lower  masses  at  lower
redshifts or  higher masses at  higher redshifts.  The curve  that best
fits the  minimum of $D_{r}$ at each redshift  (white solid line) marks a  clear distinction between the combinations  that generate a
stronger   or weaker  deflection.  This  will become  important for  the
rescaling  of the  sensitivity function,  as  we will  discuss in  more
detail in  the next  sections.   

We find  that these results  do not depend  on the specific  choice of
$M_{\rm ref}$,  with the dotted black  curve simply rescaling
vertically with  $M_{\rm ref}$. We will show in Section \ref{sec_34} how to rescale $z$ and $z_{\rm ref}$ in order to be able to compare different systems. We also
find that our results are not  significantly affected by our choice of
mass-concentration relation. The figure shows that forcing the concentration of the perturber to lie on the considered model \citep{duffy08} 
 - for which $D_r$ would be exactly zero - is not significant except 
close to $z=0$ or $z=z_s$.  A   more  detailed   discussion   on   the  impact   of   the
concentration-mass   relation   can   be   found   in     Appendix
\ref{sec_appendix_a}.

\subsection{Double lens-plane coupling with a simple
 lens} \label{ray_SIS}

We now  want to quantify the  effect of the coupling  between two lens
planes and  how much the  results of  the previous section  and Fig.~\ref{deg_nfw}  are  affected by  the  main  lens properties,  such  as
ellipticity and the presence  of an
external shear.  In order to do  so, we quantify the difference in the
deflection angle (i.e. equation  \ref{def_res}) by taking into account
the contribution  of the  main lens and  by considering  the recursive
lens  equation (\ref{2planes}).  We assume  line-of-sight haloes to be  described by a NFW  profile; at
this stage we also model substructures with NFW profiles that have the
same concentration-mass-redshift  relation as  the line-of-sight haloes,  and we
refer to Section \ref{sec_effective} for an extended discussion on the
implication of this choice.  We will discuss how to compare NFW lenses
with  substructures modeled  as  PJ  profiles in Section \ref{sec_comp_prof}.

We assume that the main lens is located at $z_{l}=0.2$ and that it is  perfectly
aligned  with  the  background  source (the  first  model  in  Table
\ref{tab_lenses}). After first modeling the main lens as a singular isothermal sphere
(SIS), we add additional complexity in the form of ellipticity (i.e.,
the main lens is a singular isothermal ellipsoid, SIE) and external
shear $\Gamma$. The external shear contributes to the deflection angle as
\begin{equation}
\alpha_{shear} = \Gamma\cdot(\cos(2\Gamma_{\theta})\cdot x+\sin(2\Gamma_{\theta})\cdot y,\sin(2\Gamma_{\theta})\cdot x-\cos(2\Gamma_{\theta})\cdot y),
\end{equation}
where  $\Gamma_{\theta}$  is the shear position angle, $(x,y)$ are the positions on the image plane relative to the centre of the main lens and $\Gamma$ is the shear strength.

For  a substructure   of  given   projected   position,  we   look  for   the
line-of-sight  halo mass  that  minimizes equation~(\ref{def_res})  at
each possible redshift.   To allow for a direct comparison, line-of-sight haloes are placed in such a way 
that they perturb the lensed images at the same projected position as substructures in the lens.  

\begin{figure}
\includegraphics[width=\hsize]{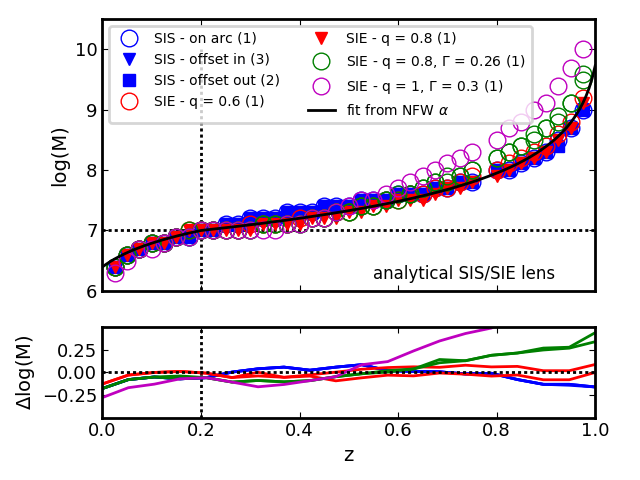}
\caption{The mass-redshift relation  for all of the considered  variations of our toy mock dataset and for a perturber with a NFW profile located at three different positions (1)-(3), corresponding to the circled numbers in Fig. \ref{lenses}. The blue symbols
represent the different  line-of-sight projected  positions in  the SIE
model that fall  exactly on the Einstein  radius of the main  lens, or
with   a  certain   offset  (following   the  numbers   from  Fig.~\ref{lenses}).  The  red   and  green  symbols   show  the
mass-redshift relation for different choices of
axial ratio for the main lens ($q$) and/or external shear ($\Gamma$). The lower panel shows the residuals for all the cases, with respect to the black curve in the upper panel. \label{deg_all}}
\end{figure}

Fig.~\ref{deg_all} shows the mass-redshift relation for different positions
of the  perturber and  different choices  of ellipticity  and external
shear strength.  The  black curve shows the best fit derived from Fig. ~\ref{deg_nfw}.  We find that for a SIS lens with no
external shear, the  results are consistent with those  derived in the
previous section  at the  5 per  cent level  and do  not significantly
depend  on   the  position  of   the  perturber,  in   agreement  with
\citet{li16b}.  For a perfectly symmetric case, the non-linear effects
arising  from  a  double-lens-plane configuration  are  therefore  not
significant.  Instead,  as we increase  the main lens  ellipticity and
the strength of the external  shear, we find stronger deviations from
the  symmetric and the single-lens-plane  cases in a way that depends on the perturber position.  
In  particular, as expected from  equation (\ref{2planes}), the deviations are stronger for background line-of-sight objects
as the deflection of the main lens enters the calculation
of the background perturber deflection  angle.

\subsection{Realistic lenses} \label{sec_34}
\begin{figure}
\includegraphics[width=\hsize]{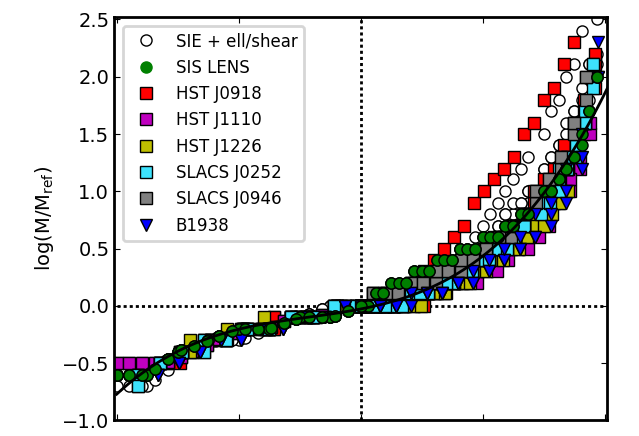}
\includegraphics[width=\hsize]{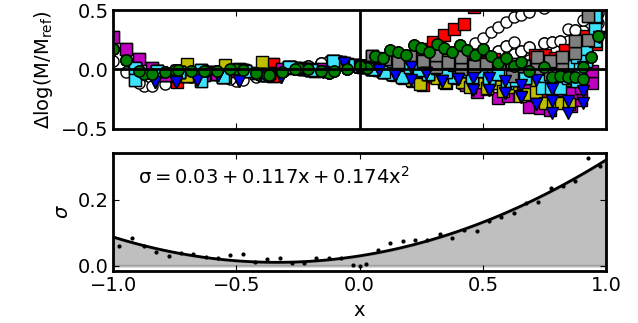}
\caption{The top panel shows the rescaled mass-redshift relation, derived by fitting all of the mock lens systems, for a NFW profile line-of-sight perturber. The black line shows the rescaled fit from equation~(\ref{univ}), while the coloured points represent the different mock lenses
used in this paper. The middle panel shows the difference with respect to the best fit, calculated as $\Delta\log(M/M_{\rm ref}=\log(M/M_{\rm ref} - \log(M_{\rm fit}/M_{\rm ref}$). The lower panel shows the scatter of the distribution of points from the middle panel around the mean; the value calculated in each redshift bin is shown by the black dots and the black line shows the best fit relation for the scatter, as given in equation (\ref{scatter}).
\label{deg_fit}}
\end{figure}

We now generalize the results  of Section \ref{ray_SIS} by considering
more  realistic lens  configurations.   Since each  lens system  among
current and future  observations has a different  combinations of lens
and  source redshift,  in  order to  combine the  results  we use  the
following   rescaled   quantities:  $y=\log(M/M_{\rm   ref})$   (where
$M_{\rm ref}$ is the virial mass of the substructure in the lens) and
\begin{equation} x =
\begin{cases} \dfrac{z}{z_{l}}-1, & \text{if}( z<z_{l}) \\
\dfrac{z-z_{l}}{z_{s}-z_{l}}, & \text{if} (z>z_{l}) \\ 0 , & \text{if}
( z=z_{l} )
\end{cases}
\label{variables}
\end{equation} 
so that $x=-1$ corresponds to the observer and $x=1$ corresponds to the source
redshift. As can be seen from Fig.~\ref{deg_fit}, this rescaling
allows us to plot all of the redshift combinations in the same parameter
space, and thus obtain a general mass-redshift relation (black solid line) 
which is given by,
\begin{equation} 
y = 0.41x +0.57x^{2}+0.9x^{3}.
\label{univ}
\end{equation} 
The best fit parameters are obtained by performing a least-squares fit
to the data points coming from the whole sample of lenses and
positions. The main panel of Fig.~\ref{deg_fit} shows the points
corresponding to all of the considered positions (numbered circles in Fig. \ref{lenses})
and lenses (listed in Table \ref{tab_lenses}), together
with the best fit curve from equation (\ref{univ}). 
The lower panels show the difference between the points and the
best-fit mass-redshift relation of equation (\ref{univ}), as well as the scatter
among the different systems. 

We  see  that  the  analytical   fit approximates  the simulated data
reasonably  well. 
When the  main lens
model includes a large external shear (as in the case of the mock data
based on BOSS J0918+5104 or the SIS+shear case), the linearity is broken and
larger  deviations  arise. In  practice,  we  find that  the  best-fit
mass-redshift relation has a scatter  that changes with redshift. The scatter is dominated by the
assumptions  made on  the  mass-concentration  relation for those
line-of-sight haloes that are in the foreground (see  Appendix
\ref{sec_appendix_a}), while for haloes  in the background, the scatter
mainly   arises    from   the    ellipticity   and    external   shear
contribution. For a fixed concentration-mass relation, the scatter among the models we consider is well 
described by the following relation,
\begin{equation}
\sigma(x)=0.03+0.117x+0.174x^{2},
\label{scatter} 
\end{equation}
where $x$ is defined as in equation (\ref{variables}).

\subsection{Comparison between different profiles} \label{sec_comp_prof}

Contrary to the previous sections  and \citet{li16b}, we now allow the
substructures  and the  line-of-sight  haloes to  have different  mass
density profiles (in particular NFW and PJ - see Section \ref{sec_mocks} for a description of the models).  
Taking into  account the possible differences  in the mass and  density profile
definitions   is    crucial  to   interpret  correctly   the
line-of-sight  contribution. Failing  to do  so may  result in   very
different   (and  incorrect) predictions for the  number   of  detectable   line-of-sight
perturbers, as we will show in the next sections.

In  this section, we  derive a relation that allows one to map the NFW virial mass into the
PJ total  mass in terms of  their relative lensing effect at  the
 same redshift. Here, the NFW profile follows the same concentration-mass relation
 as line-of-sight haloes (see Section \ref{sec_effective} for a discussion on this matter) and the properties of the PJ profile are calculated under the 
 assumption that the subhalo is located exactly on the plane of the host lens. 

Given that in all cases the projected position of the perturber needs to be close to the Einstein radius of the main lens, i.e. close to the lensed images, in order to be detected, there is no significant dependence on the projected position. Nevertheless, for a given total mass $M_{\rm tot}^{\rm  PJ}$, the PJ truncation radius still depends on the mass and the redshift of the host lens, through its Einstein radius. We used all the lenses and perturber positions from Table \ref{tab_lenses} to derive a  PJ-NFW mass conversion and test its dependence on the properties of the system.  Thus, for each lens and perturber position, we use the equations from Section \ref{secpjprof} to calculate $r_{t}$ for a set of total perturber masses and then we compare its deflection angle with that of NFW profiles via   equation (\ref{def_res}).  In general, we find that the corresponding "best fit" NFW virial mass for each PJ total mass can be calculated from the mean relation,
\begin{equation}
\log(M_{\rm vir}) = 1.07(\pm 0.1) \times \log(M_{\rm tot}^{\rm PJ})+ 0.1 (\pm 0.15), 
\label{pj_nfw}
\end{equation}
implying that  the NFW  virial mass  must be between half and one order of magnitude larger than the PJ total mass.  For the corresponding masses,  the deflection angles are similar at $r\simeq  3r_{t}$ (this value would be slightly different for a different choice of $z_{l}$ and $z_{s}$). These results parallel those of \citet{minor17} (see also Appendix \ref{sec_appendix_c}). The uncertainty on the intercept represents the scatter between the considered lens systems and reflects the fact that the PJ tidal radius depends on the mass and redshift of the host lens and the redshift of the source. In what follows we use the mean relation for simplicity. The uncertainty on the slope is related instead to the redshift evolution of the concentration-mass relation.

\begin{figure}
\includegraphics[width=\hsize]{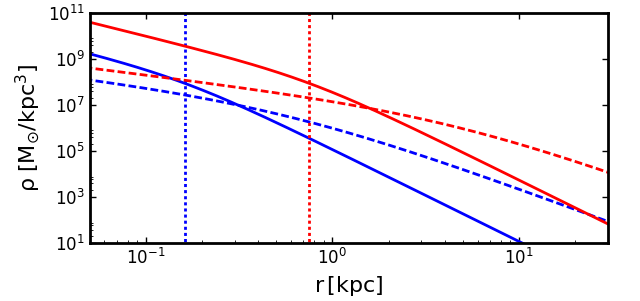}
\includegraphics[width=\hsize]{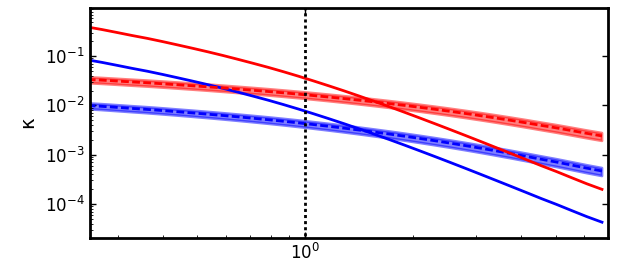}
\includegraphics[width=\hsize]{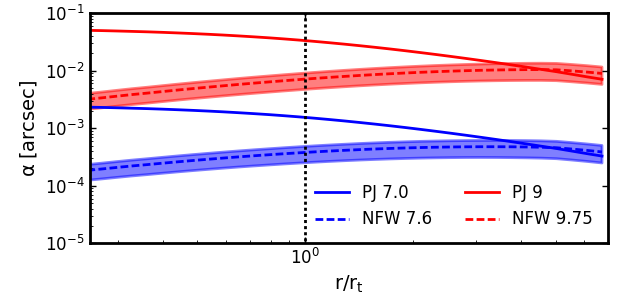}
\caption{Examples  of PJ  and NFW  masses that  give the  most similar
  lensing effect.  We show  the density  profile, the  convergence and
  the
  deflection angle as  a function of radius, for two  PJ- (solid lines)
  and  NFW-profile  (dashed lines)  masses;  the  corresponding $\log(M)$  are
  indicated in the legend of the  bottom panel and are represented by the
  same colour  and the PJ truncation  radii are marked by  the vertical
  lines of  the same colour. The  NFW halo that gives  the most similar
  effect to a certain PJ subhalo  has the same projected mass density
  within the  PJ truncation radius  $r_{t}$ and the  same deflection
  angle $\alpha$ at $\simeq 3r_{t}$.
 This also corresponds to the distance from the centre at which the
  PJ enclosed mass profile flattens, thus the enclosed mass starts approaching
  the total mass.  
  The profiles are  calculated for $(z_{\rm L},z_{\rm S}) =  (0.2,1)$, but the
  outcome is similar as a function  of redshift. The coloured bands for
  the NFW masses in the two lower panels show the result for NFW masses
  that are a factor 0.5 larger/smaller. \label{pj_radius}}
\end{figure}

For comparison, the  dashed   curves   in   Fig.~\ref{pj_radius} represent the NFW  profiles for these corresponding masses,
showing how  the enclosed mass at  $3r_{t}$ is the  same for the
two profiles.  This  relation would be  slightly different  for 
  another  choice of  concentration-mass  relation. We  can now
combine equations (\ref{univ}) and (\ref{pj_nfw}), using the latter to
rescale  the  zero-point of  the  former  and  obtain a  more  general
mass-redshift relation,
\begin{multline}
\log M_{\rm vir}(z) = \left(0.41x +0.57x^{2}+0.9x^{3}\right)+\\
\left(1.07\cdot \log(M_{\rm tot}^{\rm PJ})+ 0.1 \right)\label{univ_2}.
\end{multline}

Since this relation is equivalent to equation (\ref{univ}), modulo a  
vertical translation, it has the  same intrinsic scatter. It is
 clear from Fig.~\ref{pj_radius} that NFW profiles which follow the
field concentration-mass relation are in general not a good fit to PJ models, since both
the inner and the outer slopes are different. A better fit can be 
obtained by letting the NFW parameters $r_{s}$ 
and $\rho_{s}$ vary freely; however, this results in extremely small 
values for the former and extremely high values for the latter.  While this
would mimic the PJ profile, it would complicate the comparison between 
PJ-derived limits on substructure mass and NFW-based limits for line-of-sight haloes.  We will show in
Section \ref{sec_res2} that the mass correspondence given by equation
(\ref{pj_nfw}) is also well recovered by gravitational lens analysis of mock observations in which a PJ (NFW) perturber is  modelled using a NFW (PJ) profile. 

\subsection{The effective subhalo mass function}\label{sec_effective}

So far, we have been focusing on two definitions of mass, based on the NFW and
the PJ mass density profiles, respectively. However, the  abundance of subhaloes 
derived from numerical simulations is based on yet another mass definition. 

Subhaloes are identified in the simulations we analyse using SUBFIND \citep{springel01b,springel08b}
which locates locally overdense and self-bound regions in the density field of the host halo. The radius within which such subhaloes are  overdense is very
close to their tidal radius.  
Moreover, the properties of their density profile depends on their distance from the host centre due to stripping. Thus,  there is no reason to believe  a priori that a simulated subhalo of a given mass will produce the same lensing effect as a PJ subhalo (calculated under  the assumptions quoted in  Section \ref{secpjprof})
of the same nominal mass, or as a NFW subhalo with this mass lying on the adopted concentration-mass relation. Similarly
to what we have done in the  previous section for the comparison between PJ
subhaloes and  NFW line-of-sight haloes, we  need to make sure  that, 
for a  given $M_{\rm  low}$, we  know where to  cut the simulated subhalo mass
function on the basis of the lensing effect of the subhaloes. 
In other words, we need to rescale the subhalo mass function into an effective one,
defined in terms of an NFW profile virial mass, rather than the
subhalo mass $M_{\rm SUB}$ derived from the subhalo finder.

To this end, we use the  sample  of Early-Type-Galaxies-host  haloes with  virial masses  of $10^{13}$~M$_{\odot}$ that were  selected  from  the  Illustris  simulation   by \citet{despali17b},  and  we consider all subhaloes  with masses
$M_{\rm SUB}>10^{9}\,h^{-1}$~M$_{\odot}$ in the dark-matter-only  run.  
We fit the deflection angle of each simulated subhalo with that of a NFW profile,  and then use the NFW best-fit parameters to evaluate the corresponding
subhalo virial mass.  Unlike the previous sections, here we leave the NFW parameters free  to vary and do not impose any priors on the mass-concentration relation. 
Typically, the density profile of subhaloes - and thus their deflection angle -is well described by a NFW profile within a radius comparable to $r_{\rm max}$ - defined as the radius at which the circular velocity curve of an NFW
profile reaches its maximum value - while at larger radii the profile
is truncated.

\begin{figure}
\includegraphics[width=\hsize]{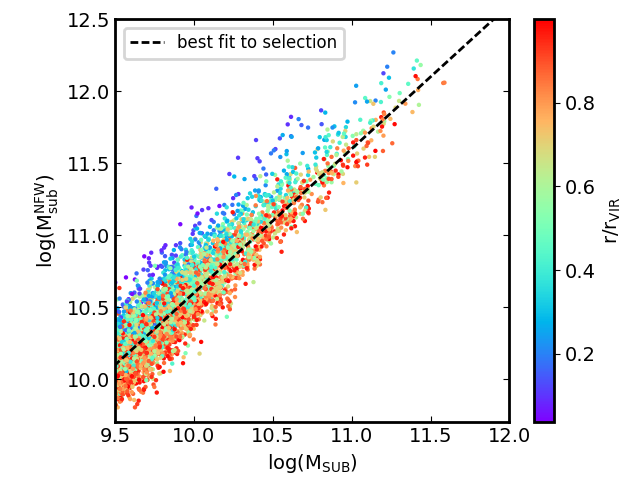}
\caption{The virial NFW mass $M_{\rm sub}^{\rm NFW}$ obtained by fitting the deflection angle of simulated subhaloes with that of NFW profiles, as a function of the original SUBFIND mass $M_{\rm SUB}$. The
points are colour-coded depending on their distance
from the centre (in units of the virial radius) of the main halo.
\label{mnfw}}
\end{figure}

Fig.~\ref{mnfw} shows  the  inferred virial  NFW masses  (hereafter
$M_{\rm sub}^{\rm  NFW}$) with  respect to  the original  SUBFIND masses
for host  haloes at $z=0.2$; the points are colour-coded according to the subhalo  distance from the centre of the host halo.  The  black-dashed line shows the  best-fit linear relation
between the  two,
\begin{equation}
\log(M_{\rm sub}^{\rm NFW})=\log(M_{\rm SUB})+0.6.
\label{submass_resc_1}
\end{equation} 
 A more precise
fitting function is obtained when the dependence on the distance
from the centre of the host halo is included, and is found to be,
\begin{equation} \label{submass_resc}
\log(M_{\rm sub}^{\rm NFW})=\log(M_{\rm SUB})+0.51-0.3\log(r/r_{\rm vir}).
\end{equation}

\begin{figure}
\includegraphics[width=\hsize]{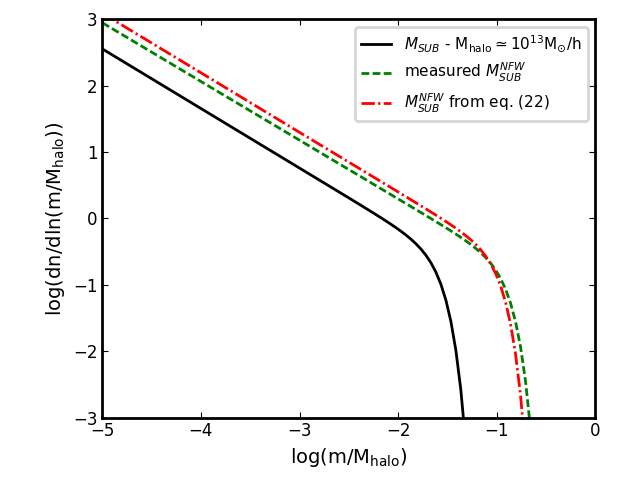}
\caption{ The rescaled subhalo mass function. 
  The black solid line shows the SUBFIND
  subhalo mass function for $10^{13}$~M$_{\odot}$ haloes at
  $z=0.2$. The green dashed line represents the new NFW virial masses
  inferred from the best fit NFW profile for each subhalo (coloured dots in Fig.~\ref{mnfw}), while the
  red dot-dashed line shows the mass function derived from rescaling the subhalo
  masses, following the best fit relation given by equation (\ref{submass_resc}), and indicated as a dashed line in Fig.~\ref{mnfw}. In
all cases, the best fit slope is consistent with $\alpha=-0.9$.
\label{mnfw2}}
\end{figure}

Fig.~\ref{mnfw2} shows the original SUBFIND subhalo mass function
(black solid line), together with the new mass function derived from the NFW fitting (green dashed line).
If we neglect the dependence on  the distance from the host centre, we
find the effective subhalo mass function to have the same slope as the
original mass  function ($\alpha=-0.9$),  but a  larger value  for the
normalization: this is shown by the red dot-dashed line in Fig.~\ref{mnfw2}.
It is important to notice that, due to this  increase  in the  normalization,  the  number of  detectable subhaloes is larger than one would derive from the original self-bound mass  function, and  thus also  larger  than what was  estimated by
\citet{li16b}.

It is worth noting that the
inferred concentration-mass relation of subhaloes differs from  the one of 
haloes in the field, that we have assumed for line-of-sight systems; the
concentrations are on average higher and are weakly dependent on the subhalo 
distance from the host centre, with higher concentrations in the innermost regions
due to the tidal truncation \citep[as already discussed among others by 
by][]{hayashi03,springel08b,moline17}. Nevertheless, we find that  $(i)$  neglecting  the  difference   in  the  concentration-mass  relation  between
  subhaloes and  field haloes leads to  an uncertainty  in the  inferred mass
  which is of order of   20 per cent for $10^{9}$~M$_{\odot}$ perturbers, and decreases with mass to $\simeq 5$~per cent at $10^{6-7}$~M$_{\odot}$ (see   Appendix  \ref{sec_appendix_c}). This translates into a shift in the total subhalo counts below 10~per cent;   $(ii)$
  neglecting changes in  the concentration with the  distance from the
  host centre translates into even smaller differences in the total
  subhalo count, to within 3~per cent. For these reasons, in what follows
  we  will assume  that  haloes and subhaloes with the same NFW  masses  have the  same
  lensing properties so that the PJ masses can be rescaled in the same
  way for both, following equation (\ref{pj_nfw}).   In this way,
  we can use the same mass limit (given by equation \ref{univ_2}) for
  both the subhalo  and the line-of-sight halo mass  function. We plan
  to study in more detail the lensing  effects and  subhalo properties
  as  a  function of  distance  from  halo  centre  in a  future  paper,
  employing higher resolution simulations. Moreover, due to the limitation in the resolution of the
observational data we are comparing with, the scales corresponding to $r_{max}$ or the PJ truncation radius $r_{t}$ are poorly resolved for low mass subhaloes. With future higher resolution observational data it may be possible to fully discriminate between the effect of different concentrations for small scale lensing perturbers.

\section{Quantifying the line-of-sight contribution}\label{sec_res1b}
In this section, we combine all of the results we obtained so far in order to quantify the line-of-sight contribution to the
  total number of detectable small mass perturbers. As mentioned earlier, the
  line-of-sight and substructure contributions can be calculated by
  integrating the halo and subhalo mass functions from the lowest
  detectable mass $M_{\rm low}$ (which is set by the observational sensitivity and angular resolution) to the
  highest possible mass for a dark clump. Here, we use
  equation (\ref{pj_nfw}) to convert the integration limit $M_{\rm
  low}=M_{\rm tot}^{\rm PJ}$ into an effective NFW mass, and then we use
  equation (\ref{univ}) to evolve the latter with redshift, and thus
  obtain an effective $M_{\rm low}(z)$ for line-of-sight
  haloes \citep[as already pointed out by ][]{Li16}. In Section \ref{sec_density}, we show how to integrate the
  (sub)halo mass function and give some examples of the total number
  of detectable (sub)haloes for specific combinations of lens and
  source redshift. To this end, we assume that the detection limit
  $M_{\rm low}$ is the same in every pixel. Then, in Section \ref{sec_sens}, 
  we show what is the effect of taking into account the full sensitivity 
  function, focusing on the particular case of SLACS~J0946+1006. The sensitivity function maps \citep[see][]{vegetti14} provide the minimum mass that can be detected for each pixel of a given system: at each position on the sky $M_{\rm low}$ can be different depending on the surface brightness of the lensed arc and other properties of the observed system.

\subsection{Integrating the mass function}\label{sec_density}

To  calculate  the number of detectable
line-of-sight haloes, we integrate the CDM halo mass function as,
\begin{equation} \label{eq_massfunc}
N_{\rm haloes}
=\sum_{i=1}^{N_{\rm pix}}\Delta\Omega_{i}\int_{0}^{z_{\rm S}}\int_{M_{\rm
  low}^{\rm NFW}(z,x_{i},y_{i})}^{M_{\rm max}^{\rm NFW}} n(m,z) \mathrm{d}m
\mathrm{\dfrac{dV}{d\Omega dz} }\mathrm{d}z,
\end{equation} 
where we  use the \citet{sheth99b} halo  mass function parametrization
and  the  best  fit  parameters appropriate  for  the  Planck  cosmology as
calculated by \citet{despali16}. Here $n(m,z)dm$ is the number of haloes
per comoving volume in the mass range $m,m+dm$. We integrate the halo mass  function in a double cone volume (as sketched in Fig.~\ref{sketch}) in order to take into account only those line-of-sight structures that may have an effect on
the  lens plane, with $\Delta\Omega_{i}$ being the solid angle corresponding to each pixel $i$. We exclude from the integration the volume within the virial radius of the host lens.  We then  obtain the  total number  of line-of-sight
haloes per arcsec$^{2}$ by dividing by the considered area in the lens
plane.  

As  discussed above, the lower integration limit $M_{low}^{\rm NFW}(z,x_{i},y_{i})$ depends on redshift and is derived from the observational $M_{low}$ using equation (\ref{univ_2}) and can vary from pixel to pixel. Since the sensitivity function (i.e the lowest detectable mass as a function of position) is different for each observed system, in this section we assume a constant limit $M_{\rm low}^{\rm NFW}(z,x_{i},y_{i}) = M_{\rm low}^{\rm NFW}(z)$, and  in the next section, we will give an example of the impact of varying $M_{\rm low}$ for each pixel. The upper  integration limit
$M_{\rm max}^{\rm NFW}$  is set equal  to $10^{11}$($10^{10}$$)$~M$_{\odot}$
for NFW (PJ) profiles. Increasing $M_{\rm max}^{\rm NFW}$ does not significantly
change our  results, due to  the exponential  cutoff at the  high mass
regime of  the halo  mass function. 

The number of detectable line-of-sight haloes calculated from equation (\ref{eq_massfunc}) has to be compared with the number of detectable subhaloes, that is, those which have a mass above the detection limit $M_{\rm low}$. In order to derive the latter, we consider the subhalo mass function of $10^{13}$~M$_{\odot}$ host haloes, as parametrized by \citet{despali17b} and rescaled as in Section \ref{sec_effective} (see also equation \ref{submass_resc}). It has been shown using simulations that the projected number density of subhaloes is roughly constant with the distance from the centre \citep{XuD15,despali17b}  for each bin in mass  and thus in order to calculate the number of detectable subhaloes we proceed as follows: we first integrate this rescaled subhalo mass function within the host halo virial radius (using the mass function parameters for the dark-matter-only case from \citet{despali17b}), and then calculate the number density of subhaloes per $arcsec^{-2}$ by dividing the total number of detectable subhaloes by the corresponding solid angle used for the integration. 

Since in  WDM models  the initial
power spectrum is suppressed below a  certain scale, the WDM halo mass
function  can  be  derived  from  the  CDM  one  using  the  relation
\citep{schneider12,lovell14},
\begin{equation} n(M)_{\rm WDM} = \left(1+\dfrac{M_{\rm cut}}{M}\right)^{\beta}
n(M)_{\rm CDM},
\end{equation}
where $M_{\rm cut}$ is  the mass associated with the scale  at which the WDM
matter power-spectrum is suppressed by 50 per cent, relative to the CDM
power  spectrum.  For  a  3.3  keV thermal relic warm  dark  matter  model, we  have
$M_{\rm cut}=1.3\times 10^{8}$~M$_{\odot}$ and  $\beta=-1.3$. The same relation
holds for  the subhalo mass  function \citep{lovell14}. 

We  remind the
reader  that we  assume  a  \citet{duffy08}
concentration-mass relation,  with the best fit  parameters for virial
masses. We  do not account  for differences in  the concentration-mass
relation between the CDM and  WDM models; as shown by \citet{ludlow16} the concentration of
WDM haloes  differs from  the CDM  case only at  low masses  (with the
exact scale, depending on the WDM  particle mass), where the number of
structures  is   also  strongly  suppressed;  finally,   lowering  the
concentration  of small  mass objects  would reduce  even further  the
number  of  detectable  (sub)haloes,  and hence  increase  the  difference
between the  two dark matter  models.  We also  stress that we  do not
consider the effect of baryons and we use the mass function taken from
dark-matter-only  simulations;  as  shown by  \citet{despali17b},  the
presence of baryons affects the number of subhaloes
in a  way that  strongly depends on  the feedback  implementation \citep[see also:][]{schaller15,fiacconi16}. The
predicted  number  of  subhaloes  reported in  this  paper  should 
therefore be interpreted as an upper limit.

\begin{figure*}
\includegraphics[width=0.48\hsize]{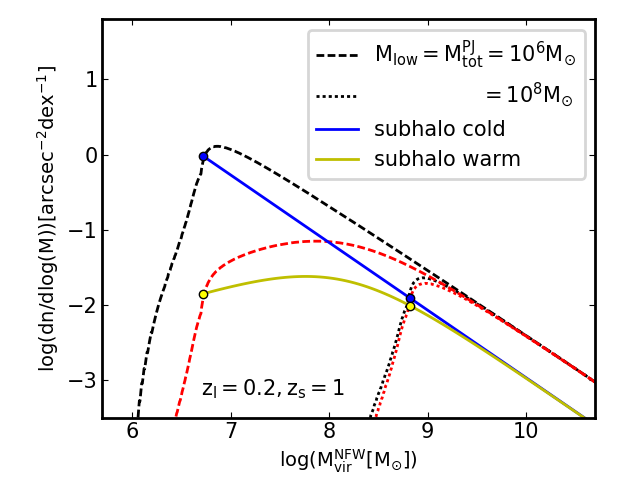}
\includegraphics[width=0.48\hsize]{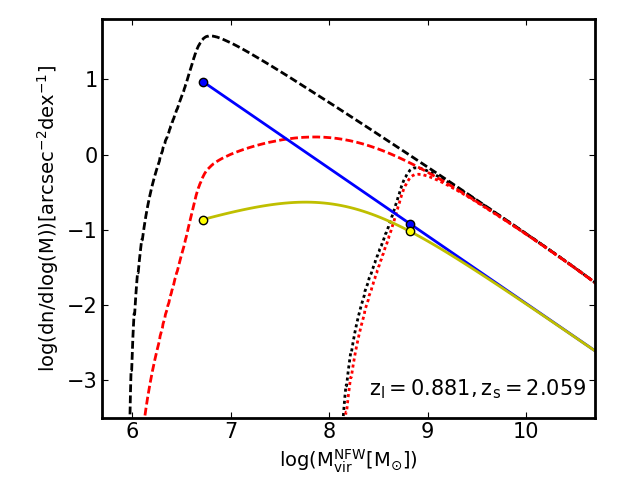}
\caption{The projected number density  of line-of-sight  haloes and subhaloes for  SLACS~J0946+1006 (left) and JVAS~B1938+666 (right).
  In       each       panel,       we       consider       a       mass
  $M_{\rm low}= M_{\rm tot}^{\rm PJ}= (10^{6},10^{8})$~M$_{\odot}$, which corresponds  to the
  minimum  subhalo mass  that can  be detected under the PJ assumption.  
Using equation (\ref{univ_2}), we  exclude from the  line-of-sight mass
function all of the  structures that cannot be  detected; the resulting
effective perturber  mass functions are  shown in all panels  by the
dashed and dotted lines. The black and red lines  represent the CDM and WDM line-of-sight mass  functions, respectively.
The blue (yellow) circles show the two values of $M_{\rm low}$ at which one should cut the subhalo mass function for the CDM (WDM) case, for the rescaled mass function and limits obtained in Section \ref{sec_effective}.  
\label{mass_func}}
\end{figure*}

In  Fig.~\ref{mass_func}  we  show the  mass function  of  line-of-sight haloes  
integrated over redshift, for two different choices of lens and source redshifts (these
 correspond  to the lowest  and highest  $z_{l}$ in our  sample) 
together with  the corresponding subhalo  mass functions, in  order to
allow  for  a direct  comparison.  In       each       panel,       we       
consider       a       mass
  $M_{\rm low}= M_{\rm tot}^{\rm PJ}= (10^{6},10^{8})$~M$_{\odot}$, which corresponds  to the
  minimum PJ subhalo mass  that can  be detected.  Using equation (\ref{univ_2}), 
  we  exclude from the  line-of-sight mass
  function all of the  structures that cannot be  detected. The resulting perturber  mass functions are calculated as
\begin{equation}
\frac{\mathrm{d} N}{\mathrm{d} \log M \mathrm{d} \Omega} = \int_0^{z_{max}} n(M,z) \frac{\mathrm{d} V}{\mathrm{d}\Omega \mathrm{d} z}   
\end{equation}   
 and are shown in all panels  by the dashed and dotted lines. In this case, $z_{max}$ is the maximum redshift at which a certain mass can be detected, calculated by inverting equation (\ref{univ_2}). In this way we calculate the total number density of (sub)haloes that can be detected for each bin in mass. The black and red lines correspond  to these CDM and WDM integrated  mass  functions respectively; the subhalo mass function is shown in blue (yellow) for the CDM (WDM) case. We find  that, increasing the lowest detectable substructure mass $M_{\rm low}$ produces
a  drastic  cut in  the  number  of observable  line-of-sight  haloes,
especially for  the CDM  case.  For a given $M_{\rm low}(z=z_{\rm L})$, the redshift-dependent cut for line-of-sight haloes has a larger impact on the number
density of  background   than  foreground  line-of-sight haloes,   since the lowest detectable mass increases rapidly in the background and the halo mass function has an exponential cut-off at the high mass end.

If instead of using equation (\ref{univ_2}), which is a median relation for all lens configurations considered in this paper, we were to use the actual relation derived for each specific case (as  presented in
Fig.~\ref{deg_fit}),  the  derived  number  density  of  detectable
line-of-sight haloes would differ at the 4 per cent level at most.

\begin{figure*}
\includegraphics[width=0.32\hsize]{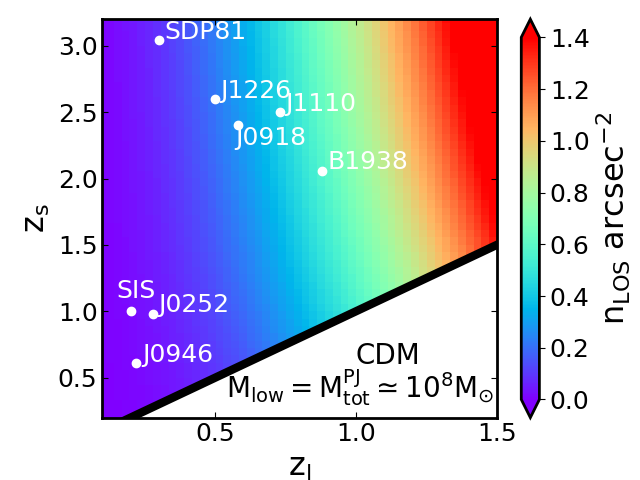}
\includegraphics[width=0.32\hsize]{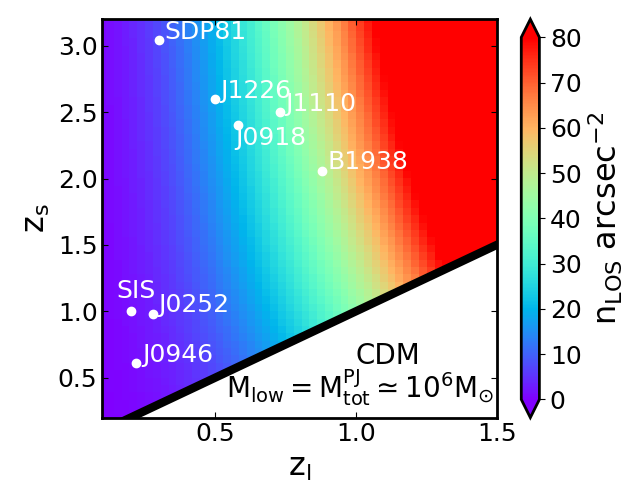}
\includegraphics[width=0.32\hsize]{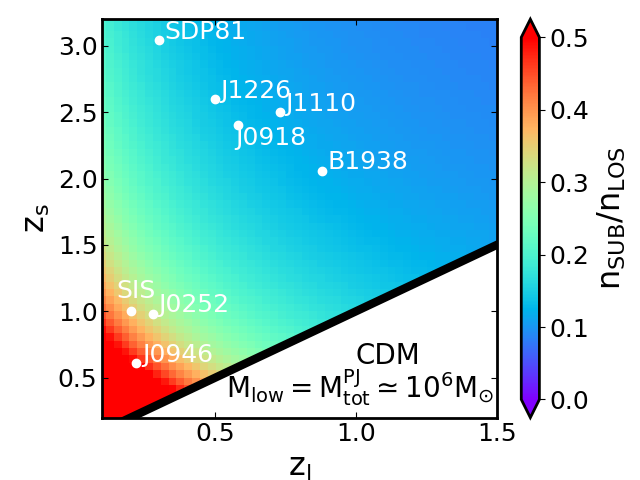}
\includegraphics[width=0.32\hsize]{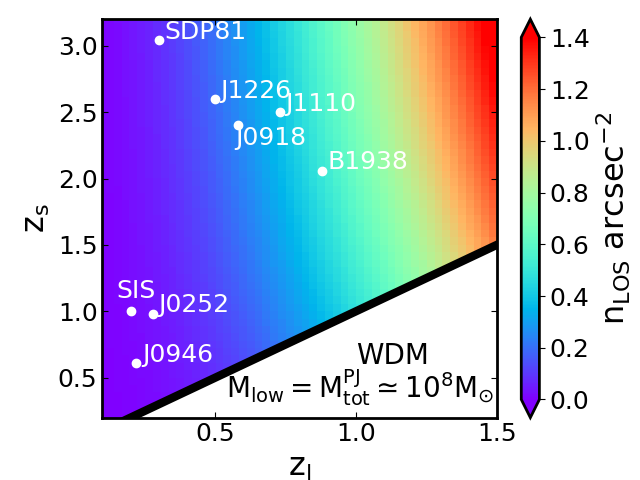}
\includegraphics[width=0.32\hsize]{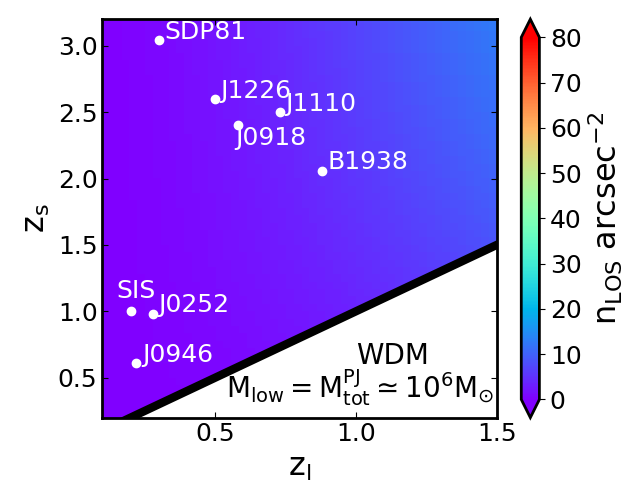}
\includegraphics[width=0.32\hsize]{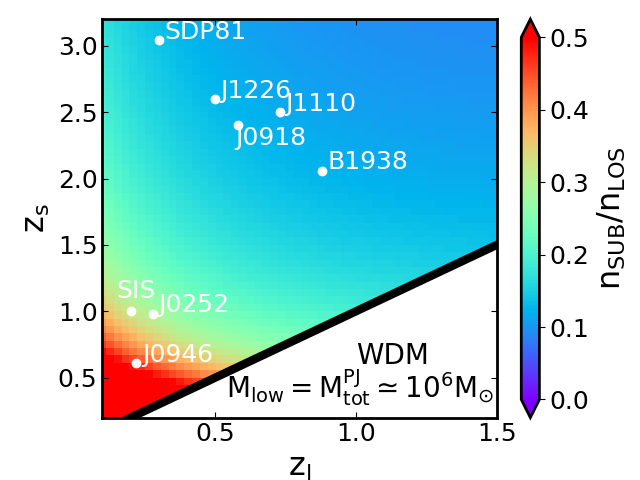}
\caption{The total number of projected line-of-sight structures per unit of arcsec$^{-2}$, for a lowest detectable mass of $10^8$~M$_{\odot}$ (left) and $10^6$~M$_{\odot}$ (middle), and for each combination of lens ($x$-axis) and source ($y$-axis) redshift. The upper panels show the results for the CDM case, while
the lower panels show the WDM case; we consider $M_{\rm low}=M_{\rm tot}^{\rm PJ}=10^{8},10^{6}$~M$_{\odot}$ (left and middle panels)
and we apply the redshift-dependent cut from equation (\ref{univ_2}) in order to calculate $M_{\rm low}(z)$ for the line-of-sight haloes.
The location in the $z_{\rm L}$--$z_{\rm S}$ plane for all of the lenses considered in this paper are marked by the white circles.  The colour-bars display the same range, both for CDM and WDM models, for each column; in the left and middle panels the color scale shows $n_{\rm LOS}$ in arcsec$^{-2}$. The fraction of detectable subhaloes with respect to the total number of  line-of-sight perturbers ($n_{\rm SUB}/n_{\rm LOS}$) is shown in the right panels for $M_{\rm low}=10^{6}$~M$_{\odot}$. As can be seen from Figure \ref{ratio_mlow} and from the values reported in Table \ref{tab_results}, the distribution of $n_{\rm SUB}/n_{\rm LOS}$ would be very similar for $M_{\rm low}=10^{8}$~M$_{\odot}$.  
\label{ratio}}
\end{figure*}

The left and central panels of Fig. \ref{ratio} (see also Table \ref{tab_results}) show the expected total projected number density of effective line-of-sight haloes $n_{\rm LOS}$ for different combinations of lens and source redshift, for two values of $M_{\rm low}$; $n_{\rm LOS}$ is expressed by the colour scale, which is the same for CDM and WDM cases, for the same $M_{\rm low}$. We notice how the two dark matter models give a similar number of predicted detectable line-of-sight haloes for $M_{\rm low}=10^{8}$~M$_{\odot}$, but how the difference  becomes striking for high sensitivity (corresponding to lower values of $M_{\rm low}$, especially when the lens and source are at high redshifts. The same conclusion can be drawn by looking at the fraction of perturbers in subhaloes (Fig.~\ref{ratio}, right panels), which is also different between CDM and WDM models, and again decreases with increasing redshift of the lens and source. The number of lenses and (non-)detections needed to discriminate between different dark matter models varies with redshift and would become smaller at high redshift, where the expected number of projected perturbers is larger.

\begin{figure}
\includegraphics[width=\hsize]{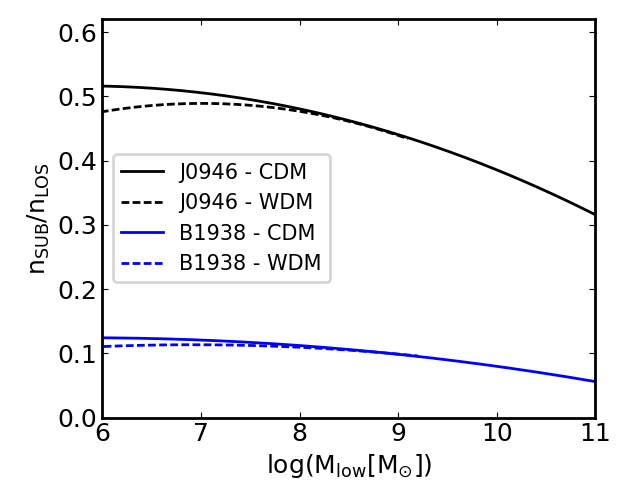}
\caption{The ratio of effective perturbers $n_{\rm SUB}/n_{\rm LOS}$ as a function of $M_{\rm low}$, for the cases of SLACS~J0946+1006 (black) and JVAS~B1938+666 (blue), both for CDM (solid lines) and WDM (dashed lines) models. \label{ratio_mlow}}
\end{figure}

One might be led to think that the value of $n_{\rm SUB}/n_{\rm LOS}$ should be independent of $M_{\rm low}$. However, this is not the case since the mass functions of the line-of-sight haloes and the subhaloes have different exponential cut-offs at the high-mass end. In Fig.~\ref{ratio_mlow}, we show how this ratio changes with $M_{\rm low}$ for the cases of SLACS~J0946+1006 and JVAS~B1938+666. The fact that the ratio is not constant indicates the importance of taking into account the variation of the lowest detectable mass from pixel to pixel. This also demonstrates the importance of increasing the angular resolution of the observational data and the discriminating power that this would bring.

From this analysis we find that the detection in the lens system SLACS~J9046+1006 by \citet[][$M_{\rm tot}^{\rm PJ}\simeq 10^{9}$~M$_{\odot}$]{vegetti10} is a true substructure with a likelihood of about 30 percent, and a line-of-sight halo with a likelihood of about 70 per cent. In general, the SLACS lenses probe a region of the $z_{\rm L}$--$z_{\rm S}$ plane in which the line-of-sight contribution is relatively limited (especially for foreground objects), since the average lens and source redshift of the sample are $z_{\rm L}\simeq 0.2$ and $z_{\rm S}\simeq 0.6$, respectively. Instead, the higher-redshift detection in the lens system JVAS~B1938+666 by \citet[][$M_{\rm tot}^{\rm PJ}\simeq 10^{8}$~M$_{\odot}$]{vegetti12} has a lower chance (below 10 per cent) of being a subhalo, and is most probably a foreground line-of-sight halo. This is also the case for the lens system SDP.81, where a substructure of mass $M_{\rm tot}^{\rm PJ}\simeq 10^{9}M_{\odot}$ has been detected by \citet{hezaveh16}.

\begin{table*} \centering
\begin{tabular}{ccccccc} 
\hline $z_{l}$ & $z_{s}$ &
$M_{\rm low}[M_{\odot}](z_{l})$ & $n_{\rm sub}$(CDM) & $n_{\rm los}$(CDM)&$n_{\rm sub}$ (WDM)& $n_{\rm los}$(WDM) \\ 
\hline 
0.2 & 1 & $10^{6}$ & 0.67 & 1.85 & 0.065 & 0.209\\
& & $10^{7}$ & 0.066 & 0.21 &0.033& 0.105\\ 
& & $10^{8}$ & 0.0063 & 0.021 &  0.006& 0.02\\ 
\hline 0.2 & 0.6 & $10^{6}$ & 0.67 &1.31&0.065 & 0.14 \\ 
& & $10^{7}$ & 0.066 & 0.15&0.033& 0.073\\ 
& & $10^{8}$ & 0.0063 & 0.016 & 0.006& 0.014\\
\hline 
0.58 & 2.403 & $10^{6}$ & 3.22 & 22.81 &0.309& 2.384\\
& & $10^{7}$ & 0.318& 2.56 &0.157& 1.235\\ 
& & $10^{8}$ & 0.030& 0.271& 0.029& 0.243\\ 
\hline 
0.881 & 2.059 & $10^{6}$ & 5.95 &46.33&  0.571& 4.482\\ 
& & $10^{7}$ & 0.587 & 5.28 &0.29& 2.41 \\ 
& & $10^{8}$ & 0.0558 & 0.57 &0.054& 0.499\\
\hline
\end{tabular}
\caption{The expected projected number density of subhaloes and line-of-sight haloes (per arcsec$^{-2}$). We count all of the (sub)haloes more massive than $M_{\rm low}$ (expressed as the lowest detectable PJ subhalo on the plane of the lens): the lower detectable subhalo mass is listed in the third column, while the corresponding value for the line-of-sight halo is calculated from equation (\ref{univ_2}). We show the results for the dark-matter-only subhalo mass function. For a WDM model, we choose the 3.3 keV thermal relic dark matter model. A generalized version of these results, spanning a wide range of both source and lens redshift, is shown in Fig.~\ref{ratio} \label{tab_results}.}
\end{table*}

\subsection{Rescaling of the sensitivity function}\label{sec_sens}
\begin{figure*}
\includegraphics[width=0.28\hsize]{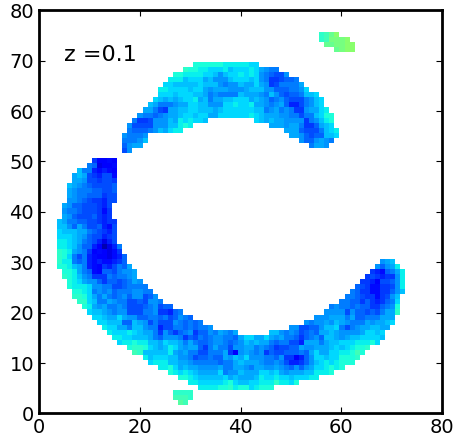}
\includegraphics[width=0.28\hsize]{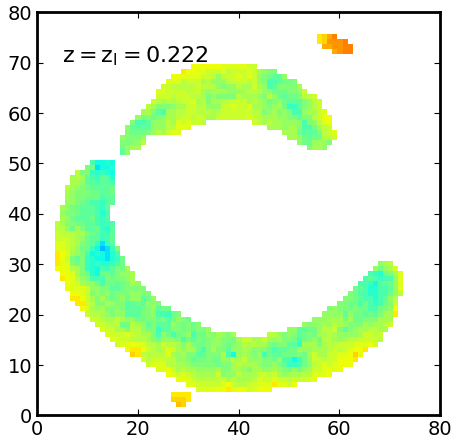} 
\includegraphics[width=0.35\hsize]{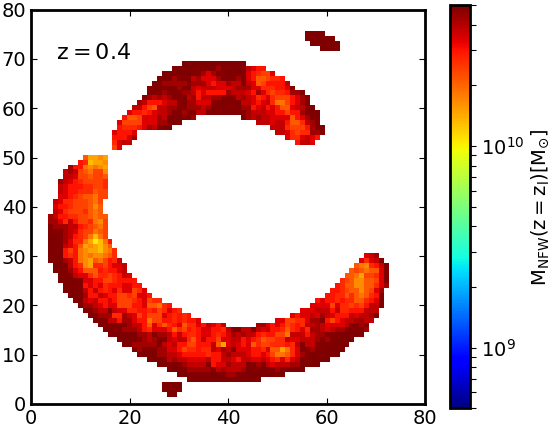}
\caption{An example of the sensitivity function for the SLACS~J0946+1006 lens. The central panel shows the lowest detectable NFW mass at the redshift of the lens ($z=z_{l}=0.222$). Using equation (\ref{univ_2}), we rescale the minimum detectable mass to redshift $z=0.1$ (left panel) and $z=0.4$ (right panel). The colour scale is the same in all panels A detailed discussion about the sensitivity function in this case, and for the other SLACS lenses, is given by \citet{vegetti14}.\label{sensitivity}} 
\end{figure*}

Here, we extend the result of the previous section to derive the total number of detectable line-of-sight haloes for one of our mock datasets and to demonstrate the role played by the sensitivity function (i.e. the smallest detectable mass as a function of projected position on the plane of the host lens). 
For this purpose, we choose  the mock dataset  based  on SLACS~J0946+1006, where a detection  of a $M_{\rm tot}^{\rm PJ}\sim10^{9}$M$_{\odot}$ subhalo was reported by \citet{vegetti10}, and for which, the full substructure sensitivity function map was
presented by \citet{vegetti14} (see also Fig.~\ref{sensitivity}).  

Using equation (\ref{univ_2}), we then derive the lowest detectable mass as a function of redshift for each pixel within a region of interest on the image plane. 
The expected number density of line-of-sight haloes is then calculated following the procedure described in Section \ref{sec_density}, with a lower integration limit for the mass function that now not only depends on the redshift, but also on the considered position according to the rescaled sensitivity function. Finally, by integrating the halo number density over the area of interest one can derive the total number of expected detections.  

In the case of SLACS~J0946+1006, we derive the expected projected number of detectable line-of-sight haloes and substructures to be 
$(N_{\rm LOS,CDM},N_{\rm LOS,WDM})= (0.036,0.035)$ and $(N_{\rm sub,CDM},N_{\rm sub,WDM})= (0.0095,0.0090)$, respectively. On the other hand, if we assume that the sensitivity is constant and equal to $2\times10^{8}$~M$_{\odot}$ (which is the lowest possible value for J0946+1006) in all pixels,  we derive $(N_{\rm LOS,CDM}, N_{\rm LOS,WDM})= (0.061,0.053)$ and $(N_{\rm sub,CDM},N_{\rm sub,WDM})= (0.025,0.023)$. This difference mainly arises because  when one considers a non-constant sensitivity function, $M_{\rm low}$ is relatively high in most pixels (at least for the realistic datasets considered here). This also results in predicted numbers that are quite similar for CDM and WDM models. Much more striking differences would arise for data with a higher sensitivity. 

\section{Mass-Redshift degeneracy}\label{sec_res2}

In  this section,  we focus  on the  mass-redshift degeneracy  between
line-of-sight haloes and substructures.  In  particular,   we  want  to  quantify  the probability that  a detection, defined  in terms of a  substructure of
measured  mass,  arises instead  from  of  a line-of-sight  halo  with a 
different mass, redshift and density profile. Our aim  is also to determine under which observational configurations  the non-linear effects arising  from the
double  lens plane  are such  that this  degeneracy can  be broken  or
alleviated. 

Note that in this Section we refer to $(x,y)$ as the position on the plane of the main lens where the perturber affects the lens images. For a subhalo this corresponds to its projected position on this plane, while for the line-of-sight haloes the latter is related to $(x,y)$ via the deflection angle of the main lens at $(x,y)$ and $\beta$ (equation \ref{beta}). 

\subsection{Modelling mock gravitational lenses}

Our  analysis thus far has been based only on analytical models and 
mock data, and to some extent did not account for any effects related to the quality or modelling of actual observational data. 
In reality, the modelling of observational data has to take into account the  signal-to-noise ratio of the images, the  effect of  the
point-spread-function (PSF), the degeneracy among the parameters of
the main  lens and of  the perturber, and  the fact that the background source is unknown and its modelled structure, which has to be inferred from  the data, can adjust to partly absorb the effects of the perturbers.

To   this  end,  we   use  the  lens  modelling   code  by
\citet{vegetti09}  to model  the realistic  lens systems  presented in
Section \ref{sec_res1}.  For each system,  the free parameters
of  the  model are  the  main  lens geometrical  parameters  (mass normalization $\kappa_0$, position,  mass density flattening $q$, position  angle $\theta$ and slope $\gamma$,
and the  external shear strength $\Gamma$ and position angle $\Gamma_{\theta}$),  the background
source  surface brightness  distribution and  regularization, and  the
perturber  mass,  projected  position  and  redshift.  As  in  Section
\ref{sec_res1},  line-of-sight  haloes  have   a  NFW  profile,  while
substructures can have either a PJ or NFW profile. 

\begin{figure*}
\includegraphics[width=\hsize]{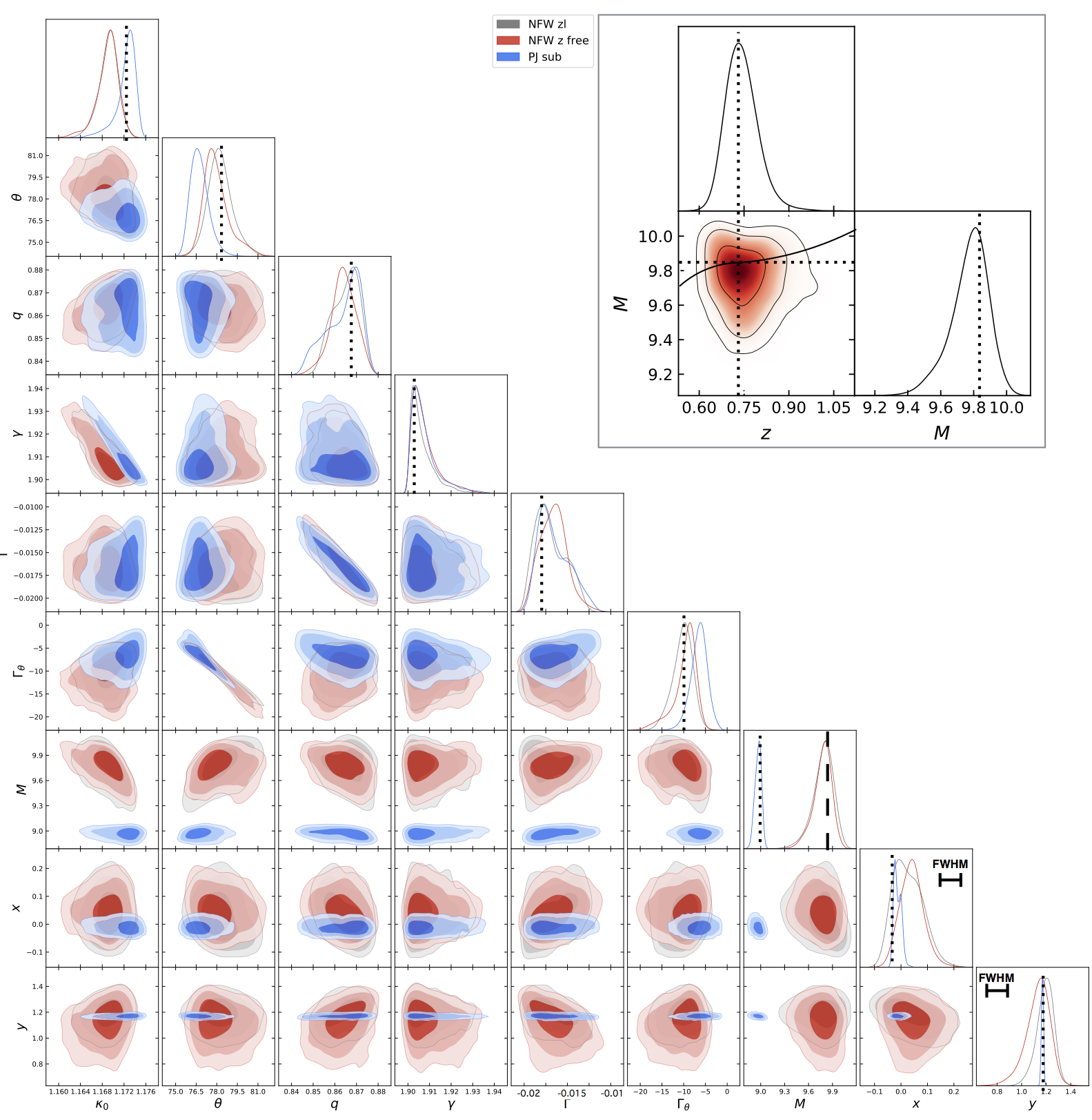}
\caption{An example of the parameter  posterior probability distributions for the case of BOSS~J1110+3649, where the mock image is created by
  adding a PJ subhalo with a mass of  $10^{9}$~M$_{\odot}$ (in position 2 from Figure \ref{lenses}).  The coloured contours  show the 1, 2
  and 3$\sigma$ levels for three different modelling choices, where we
  impose that the perturber is (i) a PJ subhalo (blue), (ii) a NFW
  subhalo (grey) and (iii)  a NFW line-of-sight halo (red), thus optimizing
  also for its redshift. The true input value for the main lens parameters are shown by the vertical dotted lines, together with the input position on the lens plane (last two columns) and the perturber mass. For this last, the vertical dotted line marks the PJ input mass, while the vertical dashed line the NFW mass predicted from equation (\ref{pj_nfw}) for this case. The two last columns also show the width of the psf FWHM for this case, in order to show that positions are well recovered. The redshift-degeneracy  for this last
  case is shown in the small inset, with the true value redshift and the predicted NFW mass marked by the dotted lines; the solid black line shows the predictions from Equation (\ref{univ_2}). As it is easy to see, the true redshift and the predicted mass are not well recovered for this case, due to its smooth surface brightness distribution and symmetry. \label{mcmc_chain}}
\end{figure*}

In  Fig.~\ref{mcmc_chain}, we  show  an  example of  the  parameter
posterior probability distributions for BOSS~J1110+3649, where the
mock image is created by adding a PJ model of a $10^{9}$~M$_{\odot}$ subhalo  at the
coordinates  $(x,y)=(0,1.15)$,  and  is  modelled  by imposing  that  the
perturber  is (i)  a PJ  subhalo (blue  contours),  (ii) a  NFW
subhalo  (grey contours),  and (iii)  a NFW  line-of-sight  halo, thus
optimizing also for  its redshift (red contours). The  last three rows
of Fig.~\ref{mcmc_chain} show  the results  for the mass  and projected
position of the perturber. The true  PJ mass is recovered for case
(i), while we infer a higher mass for cases (ii) and (iii), in
agreement  with  the expected rescaling between the NFW and PJ mass (see  equation  \ref{pj_nfw});  all of the models  recover the true perturber position well, with an uncertainty of 1--2 times the PSF full width at half maximum. The uncertainty is intended as the error with respect to the input position at the redshift of the lens, which correspond to the position of the lensing effect; a line-of-sight halo could cause a lensing effect in the same position on the image plane, even though its projected position would be different (see Figure \ref{sketch} and equation \ref{2planes}).
The constraints on the mass and redshift for case (iii) are shown in  the  inset; here, the redshift of
the lens and the NFW virial mass expected from equation (\ref{pj_nfw})
are marked  by the dotted  lines. We see that there is  effectively a
degeneracy between  the mass and redshift,  as expected, but it has a  more complicated
shape than  what is found  by comparing the deflection  angles: the black solid line shows the prediction from equation (\ref{univ_2}).  In particular,    the   uncertainty   on the redshift   
is $\Delta z\simeq  0.15$ at a $1\sigma$  level and it does  not span the
whole redshift space between the observer and the source, meaning that
not  all  the  configurations  given  by  equations  (\ref{univ})  and
(\ref{univ_2})   are equivalent. Nevertheless,
if we  force a  particular $z\neq  z_{\rm L}$ for  the NFW  perturber, the
relation from equation (\ref{univ}) still  approximates quite
well  the  recovered   mass.    

\begin{figure}
\includegraphics[width=\hsize]{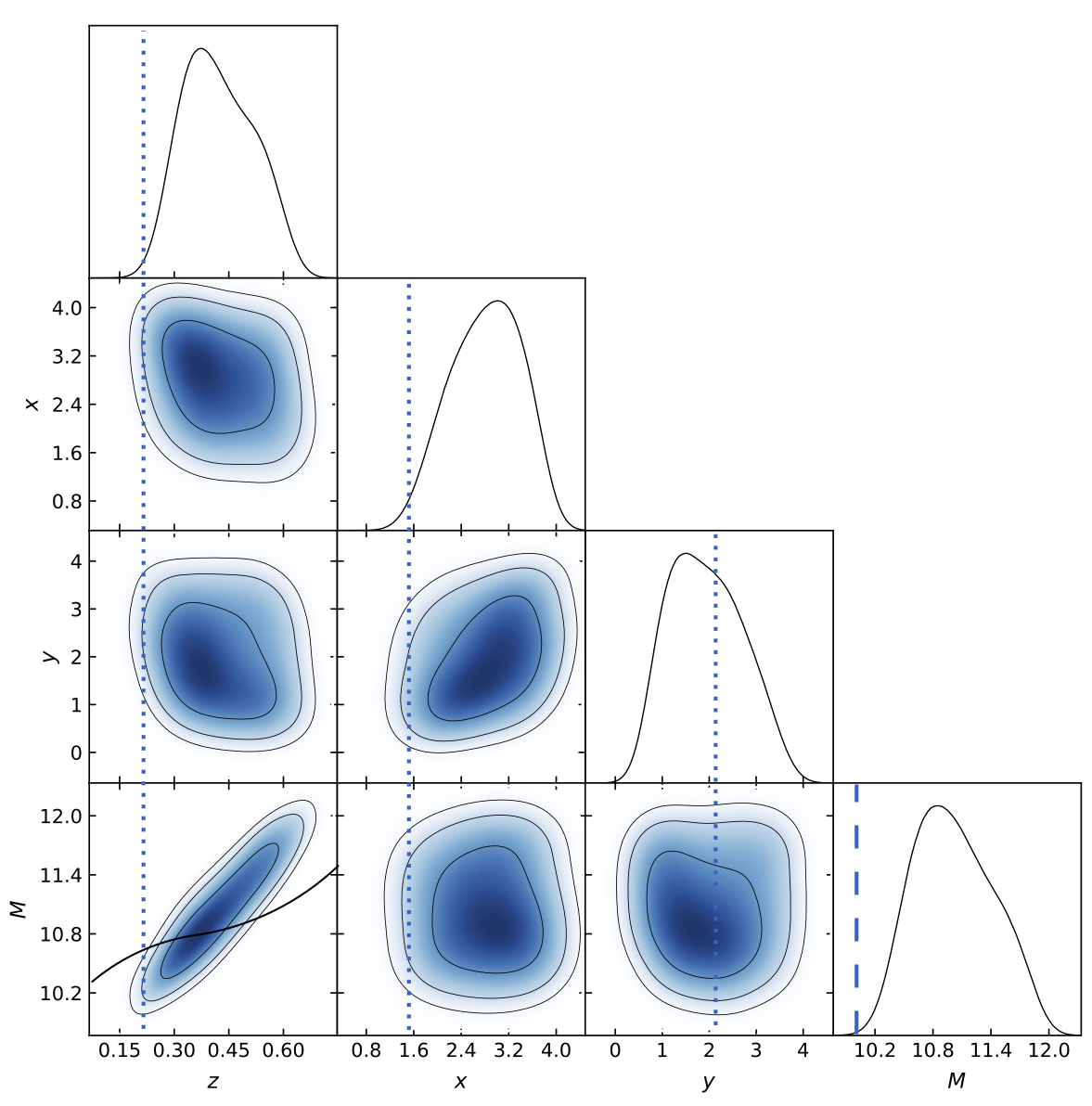}
\caption{The posterior  probability distributions  for  the  SIS mock lens.  We
  again  insert a  $10^{9}$~M$_{\odot}$  PJ subhalo when  creating the  mock
  data and we model it as a NFW line-of-sight halo. Here we show only
  the probability  contours relative  to the perturber  mass, position
  and  redshift.  The variations  in  the  main lens  parameters  with
  respect  to  the  unperturbed  model  are very  small,  due  to  the
  particularly symmetric configuration of the system. The vertical dotted  lines show the true position and redshift of the perturber; the dashed line marks the predicted NFW mass. The black curve in the $z-M$ panel  shows the predictions from Equation (\ref{univ_2}) corresponding to the peak of the posterior in mass.\label{mcmc_sis}}
\end{figure}

This happens because, using  the  image surface
brightness, and modelling the lens and source simultaneously adds
an additional  level of  information, with  respect to  the deflection
angles alone, allowing us to restrict the degeneracy range, especially for observations with a high angular resolution and a complex source surface brightness distribution. This is demonstrated in Fig.~\ref{mcmc_sis}, where  we show  the
parameter posterior  probability distributions for the  reference case
of the SIS lens at $z=0.2$;  also in this case, a $10^{9}$~M$_{\odot}$ PJ
subhalo has been  added to the lens  model and it is modeled  as in case (iii). We  see that in this  simulation the mass and redshift  are highly correlated and that the 1$\sigma$ contours span two order of magnitude in mass; moreover, even if the true position is recovered quite well
by the peak of the distribution, the uncertainties are large, spanning
almost half  of the image plane  within $3\sigma$. This is  due to the
fact that the  Einstein ring is perfectly symmetrical and the surface brightness distribution is smooth. The
width and the rounder shape of the contours also explains why for this
configuration of lens and source, the results for different  positions of  the perturber  are
equivalent (see Fig.~\ref{deg_all}). 

\begin{figure}
\includegraphics[width=\hsize]{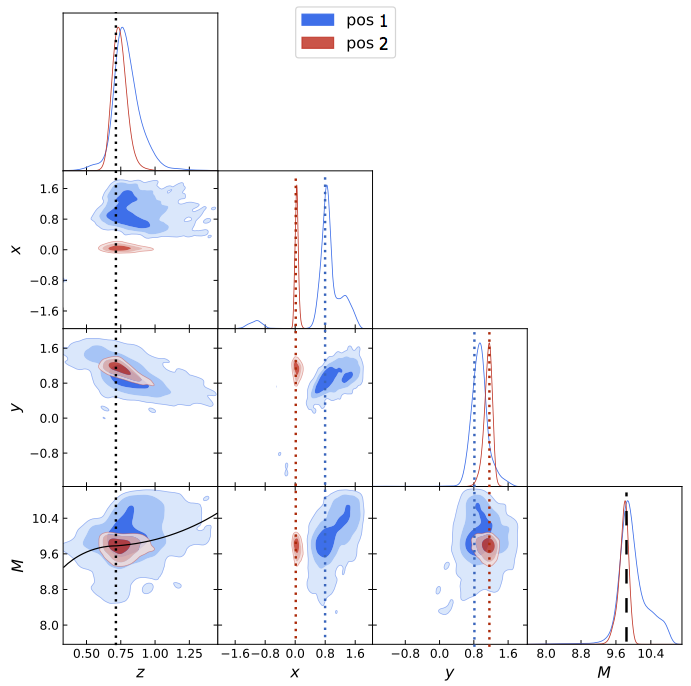}
\caption{The posterior   probability  distributions   for  two   different
  perturber positions, for the mock images based on BOSS~J1110+3649 (positions 1 and 2 from Figure \ref{lenses}). In both cases the perturber is a $10^{9}M_{\odot}$ PJ subhalo at the redshift of the lens, as in Figure \ref{mcmc_chain}. The vertical dotted lines show the input position on the plane of the lens and the input redshift; given that the true redshift (which corresponds to $z=z_{l}$) is recovered, the recovered positions are also on the lens plane. The black curve in the $z-M$ panel shows the prediction for the mass-redshift relation from equation (\ref{univ_2}) for this particular case. Finally, the dashed black line indicates the predicted NFW mass from equation (\ref{pj_nfw})}\label{mcmc_nfwz_2pos}
\end{figure}

Thus in general, the uncertainty on the mass and
redshift  depends on  the  chosen  position of  the  perturber and  in
particular the inferred quantities may  be less precise for perturbers
located  where the  surface brightness  or its  gradient is  lower: in
Fig.~\ref{mcmc_nfwz_2pos}   we  show  the  constraints   derived  by
inserting a $10^{9}$~M$_{\odot}$ PJ subhalo  at two different positions (1 and 2 from Figure \ref{lenses}) again for the case of J1110+3649, where the data signal-to-noise ratio and thus the
sensitivity    to   substructures    is   lower in position 1.     

\begin{figure}
\includegraphics[width=\hsize]{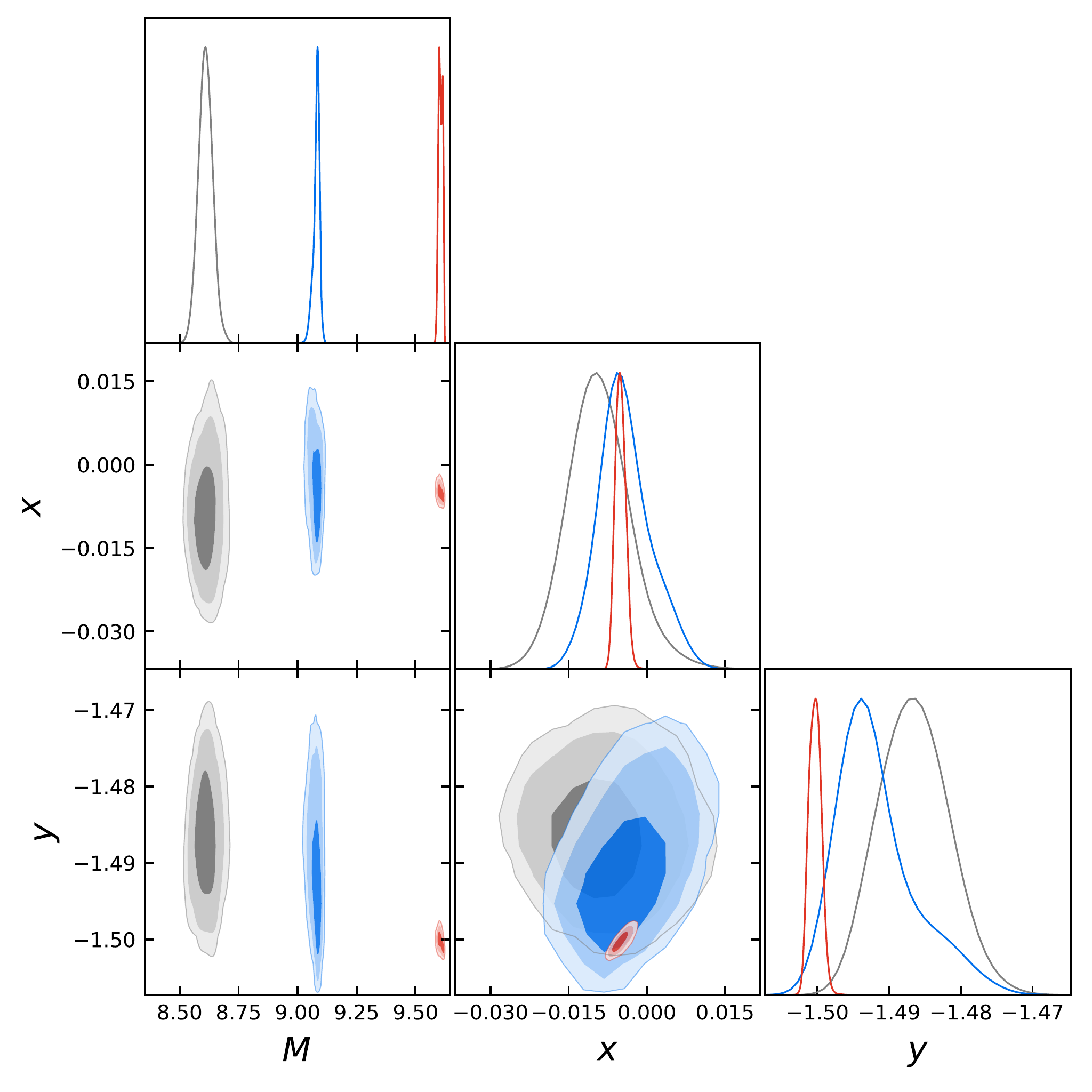}
\caption{The posterior probability distributions for SLACS~J0946+1006 
(position  1). We  see how  the  contours decrease in size and become  more
precise  as the  mass of the perturber is increased. We stress that in this plot, the whole range of $x$ and $y$ is approximately one half of the FWHM of the PSF, so the positions are very well recovered in all cases. The minimum detectable mass the pixel where the center of the perturber was placed, according to  the sensitivity function, is
$4 \times10^{8}$~M$_{\odot}$.\label{mcmc_slacs}}
\end{figure}

Finally,   Fig.~\ref{mcmc_slacs} shows the probability  contours for different subhalo
masses, all  located in the  same point, for  a system based  on
SLACS~J0946+1006. The sensitivity  function in the  chosen pixel
sets the minimum detectable mass  to be $4\times 10^{8}$~M$_{\odot}$; we see
how  the contours  are larger  when the  inserted PJ  subhalo is  only
slightly  more massive  than this  limit ($5\times  10^{8}$~M$_{\odot}$; 
grey contours), but shrink for higher mass values,  becoming more and
more precise.

Even though we ran our lensing code  on mock images for all lenses, we
only  show a  representative  subset of  contour plots.  In general, the
redshift  is  well recovered  mostly  within  $1\sigma$ and    the
corresponding  NFW virial  mass is  consistent with  the expectations from Section \ref{sec_res1}, 
even though  its exact value depends  on the exact best  fit redshift
and on the image resolution. 

In addition, from this analysis we have found that: (i) when a PJ
subhalo  is  modelled as  such,  we recover the input mass and  projected position with a precision of 0.2 dex and within $\simeq$2 FWHM of the PSF, respectively; (ii) when a PJ subhalo is modeled as
a NFW subhalo its recovered projected position is on average within 2$\times$FWHM of the PSF from the input value, while its mass is, as expected, larger than the input PJ mass. In particular, the latter differs at most by 0.4 dex from what is expected from equation (\ref{pj_nfw});  (iii) when  a PJ  subhalo is
modelled  as a  NFW line-of-sight  halo,  meaning that both the mass  and redshift   of  the   perturber  are   let   free  to   vary, the recovered redshift and projected position at which it affects the lensed images are within 0.15 and 2.5$\times$FWHM from the true values, respectively. The mass is within 0.6~$\mathrm{dex}$ from equation (\ref{pj_nfw}). (iv) When a NFW (sub)halo is modelled as a PJ subhalo, the results are consistent with the previous case, with  a reversed  ordering in  mass. 

We stress that  these are the largest uncertainties that  we have found
for realistic  lenses (thus  excluding the SIS+Gaussian  source case),
but as discussed  above, the modelling errors decrease with increasing data complexity, angular resolution and sensitivity.  In all cases,  we find that the main lens
parameters  and  source  regularization  adapt  themselves  to  partly
accommodate the presence  of the perturber and, when  necessary, for the
wrong choice  of perturber mass  profile. These  changes are at  the 3
per cent level at most, with the shear strength and the source regularization
being the most sensitive parameters.
It should be kept in mind that these results are valid for a fixed concentration-mass
relation. Moreover, these results - and in particular the ability of the code of recovering masses (PJ or the corresponding NFW) and position of the perturber with this precision - are valid for perturber masses higher than the detection threshold, defined from the sensitivity function as in \citet{vegetti14}.

\subsection{Combining line-of-sight statistics with observational constraints}

\begin{figure}
\includegraphics[width=\hsize]{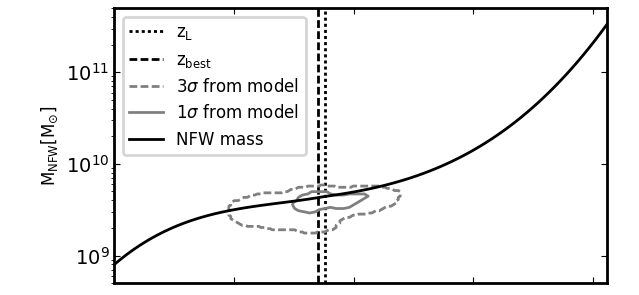}
\includegraphics[width=\hsize]{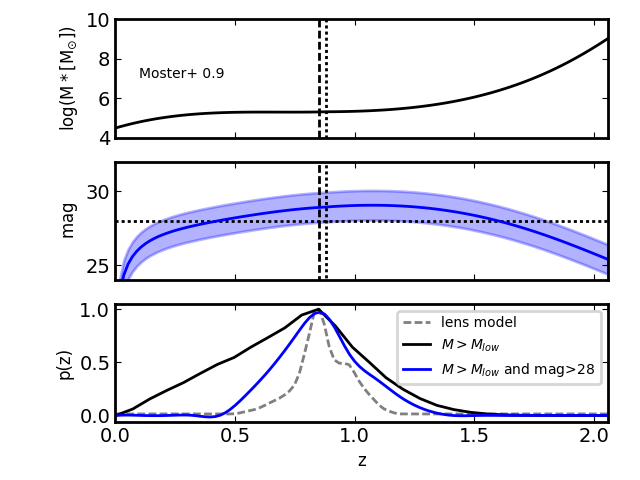}
\caption{The upper-panel shows a comparison between the predicted NFW mass as a function of redshift and that derived from the lens modelling of a mock image based on the lens model of JVAS~B1938+666. The black curve  shows the  predicted
threshold from equation  (\ref{univ_2}), while the grey contours (1 and  3$\sigma$ levels)  show the result  of the lens modelling.\label{b1938}. The upper-middle and lower middle panels show the stellar mass and corresponding apparent magnitude (with uncertainty $\pm1$ mag) associated with the perturber NFW mass from the upper panel, calculated using abundance matching arguments. The lower panel shows the redshift probability distribution of the perturber from the lens model (grey dashed line), our analytical model for the perturber mass as a function of redshift from equations (\ref{univ_2}) and (\ref{eq_massfunc}) (black solid line), and by combining our the model with the upper limit in absolute magnitude (blue solid line). In these plots, the vertical dashed line marks the redshift of the lens, while the dot-dashed line shows the redshift recovered by the lens model.}
\end{figure}

In Section \ref{sec_res1b}, we have shown how the substructure detection in JVAS~B1938+666 by \citet{vegetti12} is more likely produced by a low mass halo along the  line-of-sight to the main lens. In this section, we combine the results from Sections \ref{sec_res1} and \ref{sec_res1b} with the observational limit on the perturber magnitude ($M_{V}>-14.5$ or $K'>28$) derived by \citet{vegetti12} to set tighter constraints on the range of allowed redshifts -  under the assumption that the perturber is not a subhalo within the main lens. In particular, we first use equations (\ref{univ_2}) and (\ref{eq_massfunc}) to calculate the redshift distribution of possible line-of-sight haloes. Next, using abundance matching, we exclude all those cases for which the perturbing halo is predicted to host a galaxy brighter than the observational limit of $K'=28$. In what follows we  consider only line-of-sight haloes and exclude substructures from our calculations: thus, the virial volume of the main lens is also excluded, meaning that no line-of-sight halo can be located within the redshift range $z_{l}\pm\Delta z$ spanned by the virial radius. 

The upper panel of Fig.~\ref{b1938} shows the mass-redshift relation (black curve) derived from equation (\ref{univ_2}), together with the mass redshift degeneracy (grey contours) derived for this particular perturber (i.e. PJ substructure with a $M^{\rm PJ}_{\rm tot} = 1.8 \times 10^{8}$~M$_{\odot}$ and $z_{\rm L} = 0.881$) from the full lens modeling (see Section \ref{sec_res2}). We find that using the latter the constrains are tighter because the detailed surface brightness distribution provides additional constraining power. The two middle panels show the corresponding limit on the stellar mass and apparent magnitude \citep[derived following the formalism by][]{moster10}. The horizontal dashed line shows  the  magnitude  upper  limit  set  by  the  observational data. We can now exclude from all possible perturbers, those objects that are predicted to be brighter than $K'=28$ (most of the background objects and part of the foreground population). The effect of this selection criterion on the redshift probability distribution is shown in the lower panel. Here the grey dashed and the black solid lines show the redshift posterior probability distribution derived from the lens modeling and from our analytical analysis, respectively. In particular, the latter was defined as, 
\begin{equation}
P(z)=\frac{N_{\rm LOS}(z)}{N_{\rm LOS}({\rm tot})},
\end{equation}
where $N_{\rm LOS}({\rm tot})$ is the total number of detectable line-of-sight haloes, resulting from integrating the mass function (equ. \ref{eq_massfunc}) from $z=0$ to $z=z_{\rm S}$, and from $M_{\rm low}^{\rm NFW}$ to $M_{\rm max }^{\rm NFW}$, with $M_{\rm low}^{\rm NFW}$ derived using equation (\ref{univ_2}). $N_{\rm LOS}(z)dz$ is the total number of detectable line-of-sight haloes in the redshift range $z,z+dz$.
Finally, the solid blue curve shows the result of combining both definitions of $P(z)$ with the observational limit on the magnitude. Thanks to the inclusion of this last constraint, the redshift range can be further restricted, though most of the constraining power comes from the detailed modelling of the lensed images. In all cases, the probability peaks at or close to the redshift of the lens, meaning that the most probable location for line-of-sight pertubers is roughly within $\Delta z\simeq 0.1$ from $z_{l}$ - but outside the halo virial radius, within which they would be considered subhaloes.

\section{Conclusions}\label{sec_cloncl}
In this  paper, we  have studied the  relative gravitational  lensing effect of
substructures   and line-of-sight  haloes   on  the surface  brightness distribution of strongly lensed arcs
and  Einstein rings.   The  main  goal was  to  quantify the  relative
contribution of haloes and subhaloes to the  total number of detectable objects, as  well as to
provide  an  interpretation  of  detections  in  terms  of  these  two
populations. Our results  can be summarized as follows.

(i) Using a  set of idealized  and realistic lensing observations  we have
derived an analytic  mass-redshift relation that allows  us to rescale
the  substructure detection  threshold (i.e.  the smallest  detectable
substructure  mass)  into a  line-of-sight  detection  threshold as  a
function of redshift.  For line-of-sight haloes in the foreground with
masses much smaller than the mass of the main lens, non-linear effects
arising  from  the  double-lens-plane  configuration  are  essentially
negligible, and the above expression  provides a precise way
to quantify the abundance of detectable objects. For line-of-sight haloes
in the background of the main lens, this relation is strictly  valid only in an average
sense,  instead.  In  particular,  we find  that  departures from  the
average relations increase with  increasing asymmetries in the lensing
systems, either due to ellipticity  in the main lens mass distribution
or to the presence of a strong external shear. This translates into a small
underestimation  of   the  total   number density of   detectable  background
line-of-sight haloes of $\leq 4$ per cent.

(ii) We have highlighted  the role  played by the perturber density  
profile and  in carrying out meaningful comparison of  observations and predictions . As PJ profiles are commonly used to describe observed
subhaloes,  while modified NFW  profiles  best  describe simulated ones, we have 
derived an analytical relation that allows one to map the PJ total mass 
into the NFW virial mass based on their gravitational lensing effect. The NFW virial mass that produces the most similar lensing effect to a certain PJ subhalo (at the same redshift) is roughly one order of magnitude higher than the PJ mass: failing to correctly take into account the effect of the assumed density profile on the estimated mass can result in incorrect prediction of the expected number of (sub)haloes.

(iii) We have shown that simulated subhaloes 
can be well described by NFW haloes with a concentration-mass relation that 
is weakly dependent on the distance from the host centre and that for our purposes can be still approximated by the \citet{duffy08} relation for virial masses below $10^{9}$~M$_{\odot}$.  
Assuming a distance-independent concentration  leads 
to a small  overestimation on  the total number of detectable subhaloes 
of the order of 3 per cent.

(iv) By fitting NFW profiles to the simulated subhaloes, we have  
derived an effective rescaled subhalo mass function. This results in a 
shift of the original (SUBFIND-defined) mass function and a consequent 
increase in its normalization.  

By combining all of the above  results, we find that  line-of-sight perturbers generally dominate in number with respect to subhaloes, but  that the ratio of the two  depends strongly on the lens and source redshift due to the form of equation (\ref{univ}), as may be inferred  from simple volume arguments. For a very low detection threshold, the differences in the predicted number of detectable line-of-sight perturbers between the CDM and WDM models are particularly striking.
This reflects the fact that the abundance of WDM haloes and subhaloes is strongly suppressed relative to CDM at these masses. Future higher resolution observations with, for example, the next generation of Extremely Large Telescopes, should be able to discriminate between different dark matter models, ruling out some of them.  This ability is enhanced and made more robust by the fact that the dominant perturbing structures are expected to be line-of-sight haloes rather than substructures in the lenses. This kind of constrain on WDM can then be compared with those coming from other studies, such as works including only substructures \citep{birrer17}, focusing flux ratio anomalies \citep{inoue15} or satellite counts \citep{lovell16,lovell17}.
 
The  other main  goal of  this paper  was to  quantify the  degeneracy
between  the redshift  and the  mass of  detected perturbers. In order
to do  so, we have used the lens modelling code by \citet{vegetti09} to
analyse mock observations in which a  perturber, which may be either a
subhalo or a  line-of-sight halo, had been  artificially inserted and
modelled either  as a subhalo  or a line-of-sight halo.  The main results from this analysis are the following.

(i) The  mass-redshift relations derived from the deflection angle residuals provide a  reliable  first order estimate of the mass-redshift degeneracy. However, while equations  (\ref{univ}) and  (\ref{univ_2}),  and Fig.~\ref{deg_fit} suggested that at each redshift between the observer and the source, all masses  following the mass-redshift  relation would be  indistinguishable, we have  found instead that the  mass-redshift degeneracy is restricted  to a smaller redshift 
range that strongly depends on the complexity and angular resolution of the data.

(ii) Independent of the assumed mass density profile of the perturber, the inferred masses have a relative error of at most 0.6 dex relative to the expectation from equations (\ref{univ}) and  (\ref{univ_2}). The projected position of the perturber  is recovered to within  a few (typically 1--2) times the PSF full width at half maximum and the redshift can be constrained with  an absolute error of at most  $\Delta z\simeq 0.15$ at the  68 per cent confidence level.

We can therefore conclude that the surface brightness distribution of the lensed  images contains more information than the deflection angle, which helps to reduce the allowed parameter space, and thus improve constraints.

To  summarize,  the  contribution  from  small-mass  haloes  along  the
line-of-sight  is important  for  three reasons.  (i)  As the  lensing
effect depends on  the redshift of the  perturber,  line-of-sight
haloes that  are located  at a  lower redshift  than the  lens produce
larger perturbations  of the lensed  images than substructures  of the
same  mass inside  the lens-galaxy  halo, meaning  that the  detection
threshold is  effectively lower for foreground  objects. (ii) The
number of  detectable line-of-sight haloes is generally larger than
the  number  of  detectable subhaloes, the ratio between the two depending
on the  redshift of  the lens  and the  smallest detectable  mass.  The  line-of-sight  population
is thus an important  contribution that  significantly
boosts  the  number of  observable  small-mass  haloes and therefore tightens constraints on  the dark
matter mass  function. (iii) Line-of-sight haloes  are significantly
less affected by  baryonic processes than subhaloes, since they do not experience  significant mass loss  due to tidal  interactions. Since, in addition, they are expected in larger number, they lead to more robust and more stringent constraints  on the  properties of  dark matter.

\section*{ACKNOWLEDGMENTS}
CG acknowledges 
support from the Italian Ministry for Education, University
and Research (MIUR) through the SIR individual grant SIMCODE,
project number RBSI14P4IH. GC thanks MPA for the hospitality 
during the preparation of this work.  FvdB is supported by the Klaus Tschira 
Foundation and by the US National Science Foundation through grant AST 1516962.

\bibliographystyle{mnras}
\bibliography{paper.bbl}

\appendix

\section{Mass-concentration relation}\label{sec_appendix_a}
\begin{figure}
\includegraphics[width=\hsize]{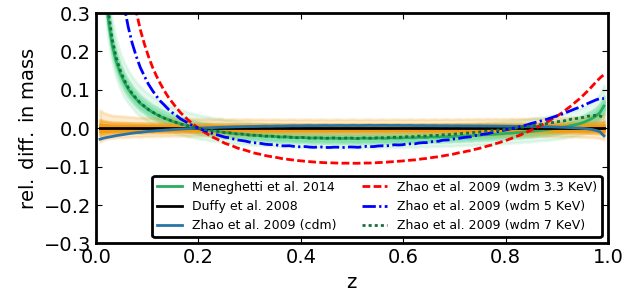}
\caption{The influence of the concentration-mass relation on the difference in the deflection angles shown in Fig.~\ref{deg_nfw}, both in CDM and WDM cases. We estimate the scatter that would be induced by using a different concentration-mass relation, comparing the models of \citet{duffy08}, \citet{meneghetti14} and \citet{zhao09}.
\label{mass_c}}
\end{figure}

In Fig.~\ref{mass_c} we plot the
relative  difference  in   the  best  fit  curve   derived  using  the mass-concentration relations by \citet{duffy08}, \citet{meneghetti14} and \citet{zhao09}, computing  the latter  both for  the CDM and  different  WDM cases (dashed  and
dot-dashed lines).  In particular, to  model the effect of  WDM within the \citet{zhao09} mass-concentration relation,  we proceed as follows: ($i$)  we modify  the  CDM  initial power  spectrum  of our  reference cosmology generated by CAMB \citep{camb} to the corresponding WDM mass as presented  by \citet{bode01},  ($ii$) we compute  the corresponding mass variances $\sigma(M)$ \citep{lacey93,sheth99b,despali16}, ($iii$)
we  adopt  the  \citet{giocoli12b}  mass accretion  history  model  to
recover the time $t_{0.04}$ at which the main halo progenitor assembled
4 per cent  of its  mass needed  by the  \citet{zhang09} concentration-mass
model. The WDM trend is opposite to the CDM one, because in WDM models
the concentration peaks  at intermediate masses and  decreases both at
the  high and  low  mass  end, behaving  similarly  to  the WDM power-spectrum \citep{ludlow16}.  The contours show the
effect of  choosing a concentration  $1$ or $2\sigma$  away from
the   average   concentration.  We find  differences that  are generally  within the  10 per cent
level, and they  become larger only towards $z=0$, where  the number of
line-of-sight haloes  is very small.  Hence, it can be  concluded that
the  specific choice  of mass-concentration  relation is  of secondary
importance.

\section{De-projection effects}
\label{sec_appendix_c}

As  detailed  in  Section  \ref{secpjprof},  the  full  shape  of  the
PJ profile depends both on its total mass $M_{\rm tot}^{\rm PJ}$ and
its truncation radius  $r_{t}$, which in turns depends  on the unknown
3D   distance  of   the  subhalo   from  the   host  centre.   From  a
substructure-lensing  modelling  point-of-view, one  generally  assumes
that the total mass and the projected distance of the substructure are
 free parameters of the model.  For a given value of $M_{\rm tot}^{\rm PJ}$
and $R$, one then derives the corresponding truncation radius under the
assumption that the  substructure is located on the plane  of the host
lens, that is, $r=R$. In practice, this implies that the value inferred for
the substructure mass is the lowest possible allowed by the data.  The
lack of knowledge  on $r$ is then taken into  account, via statistical
arguments, in  the form  of a  systematic error  on the  inferred mass
\citep{vegetti12,   vegetti14}.  This,   however,  requires assumptions  on the (controversial) substructure spatial  distribution to be made.  

In this  paper, we make use of realistic  datasets to quantify
the  error on  the  inferred total  mass that  arises  from the  $r=R$
assumption.   To  this  end,  we  create  a  set  of  mock  data  with
substructures of different total mass  and different 3D distances from
the  centre  and   model  them  using  the  lens   modelling  code  of
\citet{vegetti09}, under the assumption  that $r=R$. A similar analysis
has been recently  carried out by \citet{minor17},  however, unlike the
latter, we  do not enforce the  profiles to have the  same perturbation
scale on  the plane of  the host.  We  find that the  $r=R$ assumption
leads to a maximum error on the inferred total mass of $\simeq 85$ per
cent  for  a subhalo  located  at  the  halo  virial radius;  this  is
consistent  with   \citet{minor17}  who find  that  a
lensing  perturbation  of  the  same   scale  can  be  produced  by  a
$10^{9}$~M$_{\odot}$ subhalo that is located on the  lens plane at  the Einstein
radius  ($d\simeq 7.4$  kpc) or  by a  $8.7\times 10^{9}$~M$_{\odot}$ subhalo at
$d\simeq 100$ kpc; note that when \citet{minor17} model the  latter as a subhalo on the lens
plane  they recover  a mass  of $\simeq  1.7 \times 10^{9}$~M$_{\odot}$,
resulting in  a 79  per cent  error on  the mass  estimate due  to the
de-projection.   In  general, we  find  that  a  subhalo with  a  mass
$M_{\rm tot}^{\rm PJ}$ that is located at distance $r>R$, leads to an inferred
mass $M_{\rm inf}$ at $r=R$ given by,
\begin{multline}
M_{\rm inf}/M_{\rm tot}^{\rm PJ} = 1- 0.3 \cdot\log(r_{t,{\rm inf}}/r_{t,{\rm sub}})\\
\simeq 1- 0.3\cdot\log(r).
\end{multline}
As discussed by \citet{minor17}, one  could obtain a more precise mass
measurement by  modelling the  substructure in  terms of  their robust
mass, that is, the mass within the  distance from the subhalo centre to the
lens system critical curve along the direction where the magnification
is perturbed  the most by the  presence of the subhalo  divided by the
slope $\alpha$ of the main lens density profile. This robust radius is
larger than  the subhalo Einstein radius,  it depends on the  slope of
the main  lens mass-density profile  and is generally between  one and
two times  the PJ  truncation radius; within  this scale  the enclosed
subhalo projected  mass can be  robustly inferred, even if  the subhalo
assumed density profile and tidal radius are inaccurate.  In practice,
however,  this  mass definition  depends  on  the  slope of  the  lens
mass-density profile,  which is not  known a priori and  is degenerate
with  the  inferred size  of  the  background source.  Moreover,  this
complicates the  comparison with  the predicted subhalo  mass function
from numerical  simulations. Alternatively,  one could include  the 3D
distance as a free parameter of the model, though at present it is not
clear whether the data contains enough information to constrain it. 

What  we have shown in this  paper is that by  using a NFW profile,  one can safely ignore  de-projection effects,  as the  subhalo concentration depends only weakly on the distance from the host centre.  
We compared the lensing effect of subhaloes with different concentrations. Using mock  data  sets in  which  the  subhaloes have  a  distance-dependent concentration  and by  modelling them  with   a   NFW   profile   with   a   constant   \citet{duffy08} concentration-relation,  we  find  errors  on the  inferred  mass  and projected position respectively within 20~per cent and 3-4 pixels for perturbers of mass $10^{9}$~M$_{\odot}$. Given that the difference in the concentration-mass relation decreases with mass, the errors on the inferred mass are within 10~per cent for $10^{8}$~M$_{\odot}$ and  around 5~per cent for $10^{6-7}$~M$_{\odot}$.

\label{lastpage}
\end{document}